\documentclass[prd,preprint,superscriptaddress,nofootinbib,longbibliography]{revtex4-1}
\usepackage{graphicx}% Include figure files
\usepackage{dcolumn}% Align table columns on decimal point
\usepackage{bm,amsfonts,amsthm,amsmath,amssymb}
\usepackage{bbm}
\usepackage[]{epsfig,graphics}
\usepackage{comment}
\usepackage[T1]{fontenc}
\usepackage{lipsum}
\usepackage[utf8]{inputenc}
\usepackage{ulem}
\usepackage{multirow}
\usepackage{color}
\usepackage{lastpage}
\usepackage[sort&compress]{natbib}
\usepackage{enumerate}
\usepackage{subfigure}
\usepackage{wasysym}
\usepackage{hyperref}
\usepackage{relsize}
\usepackage{physics}
\usepackage{mathrsfs}
\usepackage{tensor}
\usepackage{xfrac}
\usepackage{siunitx}
\usepackage{setspace}  % helps with figure captions

\hypersetup{
    unicode=false,          % non-Latin characters in Acrobat?s bookmarks
    pdftoolbar=true,        % show Acrobat?s toolbar?
    pdfmenubar=true,        % show Acrobat?s menu?
    pdffitwindow=false,     % window fit to page when opened
    pdfstartview={FitH},    % fits the width of the page to the window
    pdftitle={My title},    % title
    pdfauthor={Author},     % author
    pdfsubject={Subject},   % subject of the document
    pdfcreator={Creator},   % creator of the document
    pdfproducer={Producer}, % producer of the document
    pdfkeywords={keyword1} {key2} {key3}, % list of keywords
    pdfnewwindow=true,      % links in new window
    colorlinks=false,       % false: boxed links; true: colored links
    linkcolor=red,          % color of internal links
    citecolor=green,        % color of links to bibliography
    filecolor=magenta,      % color of file links
    urlcolor=cyan           % color of external links
}

\newenvironment{itemize*}
  {\begin{itemize}
    \setlength{\itemsep}{0pt}
    \setlength{\parskip}{0pt}}
  {\end{itemize}}

\newenvironment{enumerate*}
  {\begin{enumerate}
    \setlength{\itemsep}{0pt}
    \setlength{\parskip}{0pt}}
  {\end{enumerate}}

\newenvironment{description*}
  {\begin{description}
    \setlength{\itemsep}{0pt}
    \setlength{\parskip}{0pt}}
  {\end{description}}
\def\ben{\begin{enumerate*}}
\def\een{\end{enumerate*}}
\def\bi{\begin{itemize*}}
\def\ei{\end{itemize*}}
\def\bd{\begin{description*}}
\def\ed{\end{description*}}
\def\be{\begin{equation}}
\def\ee{\end{equation}}
\def\bea{\begin{eqnarray}}
\def\eea{\end{eqnarray}}
\def\bfl{\begin{flushleft}}
\def\efl{\end{flushleft}}
\def\bigskip{\;\;\;\;\;\;\;}

\newcommand{\bcol}{\beta_{\mbox{\tiny collapse}}}
\newcommand{\bhom}{\beta_{\mbox{\tiny inhom}}}

\newcommand{\DS}{{\mbox{\tiny{DS}}}}
\newcommand{\SM}{{\mbox{\tiny{SM}}}}

\begin{document}

\title{Primordial Black Holes and Co-Decaying Dark Matter}

\author{Julian Georg}
\email{georgj9@rpi.edu}
\affiliation{Department of Physics, Applied Physics, and Astronomy
Rensselaer Polytechnic Institute, 110 8th Street, Troy, NY 12180 USA}
\author{Brandon Melcher}
\email{bsmelche@syr.edu}
\author{Scott Watson}
\email{gswatson@syr.edu}
\affiliation{Department of Physics, Syracuse University, Syracuse, NY 13244, USA}

\date{\today}

\begin{abstract}
Models of Co-Decaying dark matter lead to an early matter dominated epoch -- prior to BBN -- which results in an enhancement of the growth of dark matter substructure. If these primordial structures collapse further they can form primordial black holes providing an additional dark matter candidate. We derive the mass fraction in these black holes (which is not monochromatic) and consider observational constraints on how much of the dark matter could be comprised in these relics. We find that in many cases they can be a significant fraction of the dark matter. Interestingly, the masses of these black holes can be near the solar-mass range providing a new mechanism for producing black holes like those recently detected by LIGO. 
\end{abstract}
%\pacs{}
\maketitle
\thispagestyle{empty}
%%%%%%%%%%%%%%%%%%%%%%%%%%%%%%%
\section{Introduction}
Conventional ideas for the cosmic origin and microscopic nature of dark matter (DM) are in growing tension with observations \cite{Tanabashi:2018oca}. 
A simple and elegant DM candidate is the thermally produced weakly interacting massive particle (WIMP).  
However, after years of searches utilizing direct and indirect detection techniques, and colliders, there is still no sign of the WIMP. Moreover, little is known about the evolution of the universe prior to Big Bang Nucleosynthesis (BBN). Thermal production of dark matter requires the early universe to be radiation dominated and that the DM is in thermal equilibrium with the Standard Model (SM) -- but this need not be the case.  Indeed, fundamental theories and top-down approaches to inflationary model building suggest that other histories are possible (and often more likely) with one example being an early matter dominated epoch (EMDE) prior to BBN \cite{Kane:2015jia}. 
In addition, the requirement to achieve successful inflation implies that the inflaton must couple very weakly to other sectors.  In many cases, this implies that the SM and any hidden sectors would be gravitationally coupled at best. This suggests that hidden sectors may decouple very early from the SM and we need not expect them to be at the same temperature (see however \cite{Adshead:2016xxj}).  Establishing the expected temperature of hidden sectors is very important for restricting model building as future experiments, such as CMB-S4, will significantly improve constraints on $N_{eff}$, and such constraints rely on a knowledge of the hidden sector temperature as compared with the SM \cite{Abazajian:2016yjj}.  

Co-Decaying DM (Co-Decay) is one possible alternative to the standard WIMP paradigm. As we review below, Co-Decay \cite{Dror:2016rxc,Dror:2017gjq,Dery:2019jwf} (and also `cannibalistic' DM \cite{Pappadopulo:2016pkp,Farina:2016llk})
posits that DM decoupled in the very early universe from the SM while relativistic. The dark sector then evolves to become non-relativistic and has a temperature differing from the SM. Another interesting property is that, upon becoming non-relativistic, the dark sector particles can dominate the energy density, leading to an EMDE until one (or more) of the particles decay to the SM ensuring a radiation dominated universe prior to BBN.
As shown in \cite{Dror:2017gjq}, this EMDE can lead to enhanced DM substructure due to rapid density perturbation growth \cite{Erickcek:2011us,Fan:2014zua,Erickcek:2015bda}. 
The question we ask in this paper is: {\it can these structures further collapse to form primordial black holes (PBHs), and could PBHs be a significant fraction of the DM?}

PBHs, and whether they could be all or part of the DM, have recently received renewed interest both in 
model building \cite{Georg:2016yxa,Georg:2017mqk,Cotner:2017tir,Cotner:2016cvr,Allahverdi:2017sks,Byrnes:2018txb,Kawasaki:2018daf,Cai:2018tuh,Kohri:2018qtx,Cotner:2018vug,Allahverdi:2017sks,Gregory:2017sor,Cotner:2017tir,Carr:2017edp,Dolgov:2017aec,Domcke:2017fix,Soni:2017nlm,Quintin:2016qro,Belotsky:2018wph},
and for their observational implications \cite{Clark:2016nst,Dalianis:2018ymb,Clark:2018ghm,Takhistov:2017nmt,Guo:2017njn,Takhistov:2017bpt,Cole:2017gle,Emami:2017fiy,Inoue:2017csr,Gong:2017sie,Clark:2018ghm,Lehmann2018,Dror:2019twh}.
In \cite{Georg:2016yxa,Georg:2017mqk}, we considered whether PBHs could form in the presence 
of string moduli that lead to an EMDE, and whether they could be a significant component of the DM. An important difference of PBH formation in these models is that, 
unlike many approaches 
to PBH formation that predict a monochromatic spectrum, the extended EMDE leads to continual production of PBHs over a range of masses. For the EMDE of \cite{Georg:2017mqk}, CMB constraints on the lightest PBHs then forced the total abundance of PBH DM to be negligible.  

In this paper, we consider PBH formation in the Co-Decay scenario, where the duration of the EMDE is expected to be much shorter lived than that of \cite{Georg:2017mqk}.  This will weaken some of the constraints on the PBH formation process leading to the possibility of PBHs making up a significant fraction of the DM. 
The rest of the paper is as follows. In the next section, we review Co-Decay focusing on the aspects relevant for PBH and DM production. In Section \ref{pbh}, we provide a somewhat detailed discussion of PBH production in an EMDE closely following the early work of \cite{Nov1975,Zeldovich:1969sb,Zel:1983,Dorosh1970,Dorosh1978,KP81}. In Section \ref{cdpbh}, 
we present our main results giving the extended mass function of PBHs from Co-Decay and placing constraints on their DM abundance using observations. We provide our conclusions in the final section. 

\section{Co-Decaying Dark Matter}
\begin{figure}
\begin{center}
\includegraphics[height=1.9in]{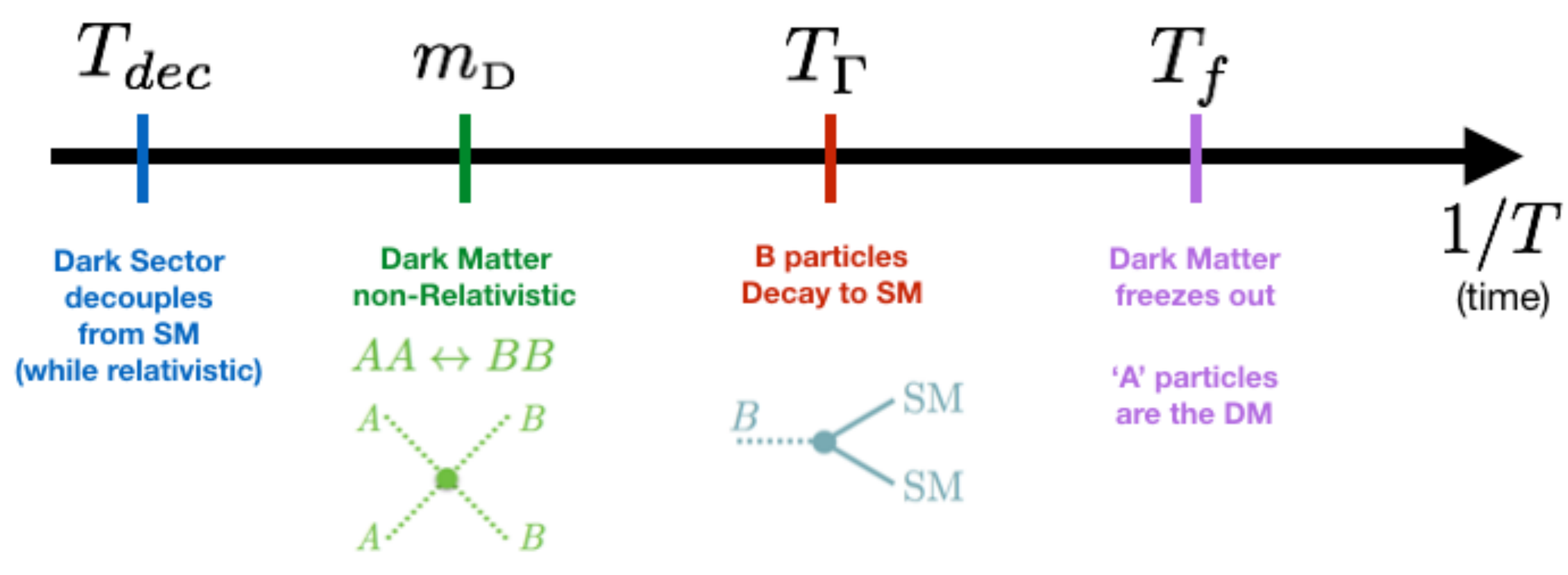}
\end{center}
\caption{Timeline for Co-Decaying dark matter. Hidden sector particles decouple very early from the SM and while relativistic. Hidden sector interactions then keep the dark abundances near equilibrium at a temperature that is typically different than the SM. In the simplest models, one particle decays to the SM, while the other is stable providing a dark matter candidate.}
\label{cddm}
\end{figure} 
Co-Decaying Dark Matter \cite{Dror:2016rxc,Dror:2017gjq} posits the existence of a dark sector with at least two particles that decoupled from the SM in the very early universe.  In the simplest model with only two particles, one is stable -- providing a possible dark matter candidate -- whereas the second can decay to the SM. For simplicity we can take these dark sector particles to be nearly mass degenerate and we label them $A$ (which plays the role of DM) and $B$ (which decays to the SM), and denote their mass as $m$.  

At some early point in the cosmological evolution, the dark sector and SM fall out of equilibrium with each other.  Until the $B$ particle decays with rate $\Gamma_{\mbox{\tiny{B}}}$, there will be negligible entropy transfer between the dark sector and SM.  Therefore, each sector's entropy is approximately conserved until that point. It will be useful below to consider the ratio of the entropy densities of the dark sector and SM at decoupling,
\be
\xi\equiv \frac{s_{\mbox{\tiny{DS}}}}{s_{\mbox{\tiny{SM}}}}\sim\mathcal{O}(10^{-1}).
\ee
After decoupling, $AA\leftrightarrow BB$ interactions\footnote{\cite{Dror:2016rxc} also considered the possibility of number changing processes between the $A$ and $B$ particles.  While this does affect the relic abundance for the DM, it does not affect the duration of the matter domination and this duration is the main relevant quantity for the consideration of PBH formation, as we will see below.  For this reason, we neglect the effect of `cannibalism' for the rest of the paper.} (with threshold s-wave annihilation rate denoted $\sigma$) maintain equilibrium between the $A$ and $B$ number density.  When the temperature of the dark sector becomes order of the mass of the dark sector particles $T_{\mbox{\tiny{DS}}}\sim m$, they become non-relativistic and scale cosmologically as a pressure-less gas ($p=0$).  In many cases the dark sector particles have an initial abundance that after becoming non-relativistic will naturally lead to an EMDE prior to BBN \cite{Dror:2017gjq}.  Once the dark sector particles become non-relativistic $T_{\DS} < m$ we have  
\be \label{equality}
\rho_{\DS} = m s_{\DS} = m \xi s_{\mbox{\tiny{SM}}} = m \xi \frac{\rho_{\mbox{\tiny{SM}}}}{T_{\SM}},
\ee
\begin{figure}[t!]
\begin{center}
\includegraphics[scale=0.8]{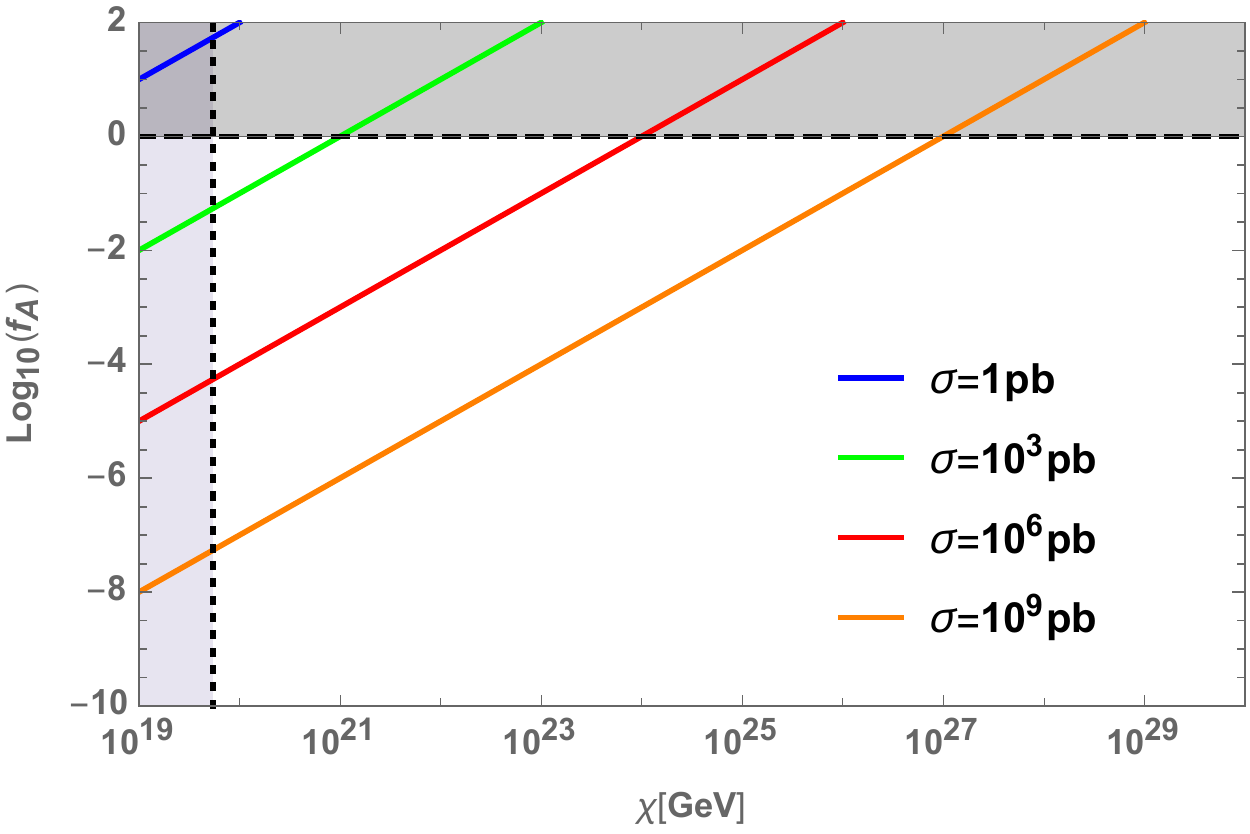}
\end{center}
\caption{The relic abundance of the A particles in the Co-Decay scenario as a function of $\chi \equiv m_{\mbox{\tiny{A}}}^2/\Gamma_{\mbox{\tiny{B}}}$ for a variety of threshold cross-sections.  The vertical, dotted, black line indicates the value of $\chi$ for which $N_{\mbox{\tiny dur}}=0$ (there is no EMDE). The horizontal dashed black line indicates the point at which the A particles can fully account for the observed present-day dark matter abundance.}
\label{Abund}
\end{figure} 
which implies that dark sector-SM equality corresponds to $T_{\SM}=T_{eq}=\xi m$.  If an EMDE begins, it will last until the time at which the $B$ particles begin decaying to the SM, $H^{-1}>\Gamma_{\mbox{\tiny{B}}}^{-1}$.  The duration of the EMDE is conveniently expressed in terms of e-folds, $N = \log a$.  Assuming instantaneous decay, the duration is given by
\be
N_{\mbox{\tiny dur}} = \log a_\Gamma - \log a_{\mbox{\tiny{MD}}} = \frac{1}{3} \log \left( \frac{s_{\mbox{\tiny{SM,MD}}}}{s_{\mbox{\tiny{SM,$\Gamma$}}}} \right), 
\ee
where MD denotes the onset of the EMDE, and the final equality follows from conservation of co-moving entropy\footnote{In practice, entropy production will occur continuously throughout the EMDE as discussed in \cite{Dror:2017gjq}, however using the instantaneous approximation here will not change our main conclusions.}.  
At the end of the EMDE, we have that $H \sim \Gamma_{\mbox{\tiny{B}}}$ and we find the number of e-folds 
\be
s_{\mbox{\tiny{SM,$\Gamma$}}} \simeq \frac{\Gamma_{\mbox{\tiny{B}}}^2 m_p^2}{m \xi}, \quad s_{\mbox{\tiny{SM,MD}}}\simeq (m\xi)^3\Rightarrow N_{\mbox{\tiny dur}} \simeq \frac{2}{3} \log \left( \frac{m_{\mbox{\tiny{A}}}^2 \xi^2}{m_p \Gamma_{\mbox{\tiny{B}}}} \right),
\ee
where we also used $s_{\mbox{\tiny{SM}}}\sim T^3$.  For reasonable choices of parameters, the EMDE will lead to $0\leq N_{\mbox{\tiny dur}} \leq 20$ \cite{Dror:2017gjq}.  

Due to the dark sector interactions, the $A$ particle number density will also decrease with the decaying particle's number density until dark sector interactions become negligible, which follows from their velocity-averaged cross-section.  To obtain the present day relic density, one must calculate the number density of the $A$ particles at this freeze-out point $H \approx n_{\mbox{\tiny{A}}}\langle\sigma v\rangle$, and then evolve it forward to the current time.  The relic abundance predicted by Co-Decay is then \cite{Dror:2016rxc}
\be
\label{aAbund}
f_A\equiv\frac{\Omega_{A}}{\Omega_{DM}} = \left( \frac{1 \; \si{pb}}{\sigma} \right) \left( \frac{m_{\mbox{\tiny{A}}}}{1 \;\si{GeV}} \right)^2 \left( \frac{10^{-18} \; \si{GeV}}{\Gamma_{\mbox{\tiny{B}}}} \right).
\ee
Co-Decay abundance can match the observed DM relic abundance for a large portion of parameter space and is plotted in Fig. \ref{Abund}.

We have seen that Co-Decay can induce an EMDE.  In a matter dominated universe, perturbations grow linearly with the scale factor, and the same result was shown to hold in the 
Co-Decay context \cite{Dror:2017gjq}.  In that paper, it was found that sub-structures could grow during the EMDE. Given that the dark matter decoupled from the SM far before the EMDE, this can lead to very concentrated substructures that may survive until today and lead to boosted signals for indirect detection experiments. Instead, here we want to see whether non-linear growth of these structures can lead to the formation of PBHs.  In the next section, we review the formation of PBHs in a matter epoch, and then we estimate the size and amount of PBHs we expect to form. 

\section{PBH Formation in an Early Matter Phase \label{pbh} }
The density contrast on sub-Hubble scales grows as the scale factor $\delta \equiv \delta \rho / \rho \sim a(t)$ during the EMDE driven by Co-Decay \cite{Dror:2017gjq}. 
Once these fluctuations grow to $\delta \sim {\cal O}(1)$ linear perturbation theory fails and one must turn to other methods. In this paper, we will use the Zel'dovich Approximation \cite{Zeldovich:1969sb,Zel:1983} to estimate the abundance of PBHs.  This method has been used by a number of authors to estimate PBH production in an EMDE \cite{Nov1975,Zeldovich:1969sb,Zel:1983,Dorosh1970,Dorosh1978,KP81} -- and more recently in \cite{Kokubu2018,Harada2016,Georg:2017mqk,Georg:2016yxa,Malec2015}. The approximation has been shown to be in good agreement with full N-body simulations until the time of shell-crossing (caustic formation) \cite{coles} and we leave a full N-body simulation to future work. The majority of this section can be understood by piecing together various parts of the references mentioned above, but here we want to present a self-contained, but brief review, and establish our notation for the next section. 

\subsection{The Zel'dovich Approximation \label{Zapprox}}
The Zel'dovich Approximation makes investigating the non-linear regime of density perturbations feasible by doing perturbation theory in the distance a particle moves from its initial value as opposed to requiring $\delta \ll 1$. Given that the perturbations have their initial values set by inflation, we expect the particles to be 
nearly homogeneous and isotropic (also in good agreement with observations). 
Then, as the perturbations enter the Hubble radius and eventually the non-linear regime, one can estimate the probability of PBH formation by establishing whether the collapse preserves enough sphericity to collapse to a PBH.
To make this quantitative, we can write the physical location of the particles separated into their 
background value and the separation resulting from evolution of the perturbations, 
\begin{equation} \label{lcoord}
\vec{r}(t,\vec{q})= a(t) \left[ \vec{q} -b(t) \nabla_q \Phi(\vec{q})  \right],
\end{equation}
where $a(t)$ is the scale factor so that $\vec{r}=a(t) \vec{x}$ are the physical coordinates, and the co-moving coordinate $\vec{x}$ is split into its initial location $\vec{q}$ (Lagrangian coordinate) and the change resulting from the perturbations -- where we again emphasize we are taking the particle separation to be small, not the density perturbation.  

Initially, when a mode enters the Hubble radius we have $b(t_H)=0$, and then the time dependence of the fluctuation leads to particle separation. This will evolve in the linear regime as
\be
\ddot{b}+2H\dot{b}=4\pi G \rho \, b,
\ee
which is the same as the equation for the linearized density perturbation in a pressure-less universe.  Therefore, we have $b(t) \sim a(t) \sim t^{2/3}$ for the time dependent growth in the linear regime. The other term in \eqref{lcoord} is related to the velocity perturbation as
\be \label{irrotate}
\delta \vec{v} \equiv \frac{d \vec{r}}{dt}-H\vec{r}=a \frac{d\vec{x}}{dt}=-a\dot{b}\nabla_q \Phi,
\ee
implying that the velocity perturbation can be written as a gradient, i.e. the fluid is irrotational. 
Using the Poisson equation (see e.g. \cite{Brandenberger:2003vk}) one can then show $\delta(t,\vec{r})= b(t) \nabla^2_q \Phi(q)$.

We now want to consider the non-linear regime ($\delta \sim {\cal O}(1)$). For small particle displacements in \eqref{lcoord} we can define a linear map between $\vec{q}$ and $\vec{r}$ allowing us to relate the energy density at later times to the initial average energy density as
\be \label{getenergy}
\rho(t,\vec{r})= \frac{\langle \rho(t) \rangle}{\det \left( D \right)},
\ee
where
\be \label{deftensor}
J_{ij}=\frac{\partial r_i}{\partial q_j}=a(t) \delta_{ij} -a(t)b(t) \frac{\partial^2 \Phi}{\partial q_i \partial q_j},
\ee
where $J$ is the Jacobian of the coordinate transformation and the shift in particles trajectories, $D_{ij}=J_{ij}-a(t)\delta_{ij}$, is the deformation tensor.
Given the irrotational flow, the Jacobian is symmetric and diagonalizable, and
we can express \eqref{getenergy} in terms of the eigenvalues as
\be \label{getmoreenergy}
\rho(t,\vec{r})= \frac{\langle \rho(t) \rangle}{\left(1+b(t)\alpha\right)\left(1+b(t)\beta\right)\left(1+b(t)\gamma\right)},
\ee
where $\alpha,\beta,$ and $\gamma$ are the eigenvalues of the deformation tensor.
It follows from \eqref{getmoreenergy} that in the linear regime 
\be
\delta(t,\vec{x}) =\frac{\delta\rho}{\bar{\rho}}\simeq - \left( \alpha + \beta + \gamma \right) b(t),
\ee
as expected.  In general, negative eigenvalues correspond to a direction collapsing.  If there is more than one negative eigenvalue, the more negative one will imply the corresponding direction collapses faster. To get PHB formation, we need the eigenvalues to be roughly the same (symmetric, spherical collapse), and we will see that this requirement leads to a suppression in the formation rate since it corresponds to a less likely configuration.  Also, if we consider one negative eigenvalue (say $\alpha$) in \eqref{getmoreenergy}, at the time $b(t_c)=-\alpha^{-1}$ the energy density is infinite.  This is when a caustic forms (particle locations intersect), and then \eqref{deftensor} can not be diagonalized.  We will next consider the distribution of the eigenvalues in the deformation tensor, which will be important for addressing caustics and sufficiently spherical collapse.

\subsection{Probability Distribution Function of the Eigenvalues}
As discussed above, the deformation tensor, $D_{ij}$, is a symmetric matrix with six entries contained in the Jacobian \eqref{deftensor}.
The distribution of its entries were 
calculated some time ago by Doroshkevich \cite{Dorosh1970} assuming the fluctuations to be random, be Gaussian-distributed, and have zero mean.

The multivariate Gaussian distribution function of the six independent entries can be written
\begin{equation}
\label{PDF1}
w(\varphi_1\dots\varphi_6)\dd{\varphi_{1\dots6}}=\frac{|C^{-1}_{pq}|^{\frac12}\exp\left [-\frac12 \sum_{p,q}\varphi_p C^{-1}_{pq}\varphi_q \right ]}{(2\pi)^3}\dd{\varphi_{1\dots6}}
\end{equation}
where we have defined $\varphi_{1,2,3}=D_{ii}$ for $i=1,2,3$ (diagonal components) and $\varphi_{4,5,6}=D_{ij}=D_{ji}$ for $i\neq j$ (off-diagonal components). The covariance matrix $C_{pq}$ gives the correlation between the deformation tensor elements
\begin{equation}\label{eq:coma}
C_{pq}\equiv\expval{\varphi_p\varphi_q}=\expval{D_{ij}D_{kl}}=\frac{\sigma^2}{15}(\delta_{ij}\delta_{kl}+\delta_{ik}\delta_{jl}+\delta_{il}\delta_{jk}),
\end{equation}
where
%\begin{equation}
%\sigma^2=\frac{1}{2\pi^2}\int \mathcal{P}(k)k^2\dd{k},
%\end{equation}
$\sigma^2$ is the variance and $C_{pq}$ is a $6\times6$ and block-diagonal matrix:
\be
C= \frac{\sigma^2}{15}
\begin{pmatrix}
A & 0 \\
0 & \mathbbm{1}
\end{pmatrix},  \quad \mathrm{with}  \quad 
A=
\begin{pmatrix}
3 & 1 & 1 \\
1 & 3 & 1 \\
1 & 1 & 3
\end{pmatrix}.
\ee
Using this expression for $C_{pq}$ in \eqref{PDF1} gives the probability distribution function (PDF) for the values of the deformation tensor
\begin{equation}
\label{eq:pdf}
w(D_{ij})\dd[6]{D_{ij}}=\frac{675\sqrt{5}}{16\pi^3\sigma^6}\exp[N_D]\dd[6]{D_{ij}},
\end{equation}
with
\begin{equation}\label{eq:exp}
N_D=-\frac{3}{\sigma^2}\left [ (D_{11}^2+D_{22}^2+D_{33}^2)-\frac 12(D_{11}D_{22}+D_{11}D_{33}+D_{22}D_{33})+\frac{5}{2}(D_{12}^2+D_{13}^2+D_{23}^2)\right ].
\end{equation}

Again, recalling that the Jacobian \eqref{deftensor} is diagonalizable, and therefore so is the deformation tensor, we can diagonalize it
through an $SO(3)$ rotation $D=R^T \lambda R$ to find the six unknown elements in terms of the eigenvalues $\lambda=\mbox{diag}(\alpha, \beta, \gamma)$.
The elements of the matrix $R_{ij}$ are the directional cosines of the new coordinate system with respect to the old one. The angles specifying the rotation are the Euler angles $\theta$, $\psi$, $\varphi$ (see, e.g. \cite{goldstein2002classical}) and 
in this new coordinate system the PDF \eqref{eq:pdf} becomes
\begin{equation}
w(D_{ij})\dd[6]{D_{ij}}=w(\alpha,\beta,\gamma,\theta,\psi,\varphi) \left | \frac{\partial^6 D_{ij}}{\partial (\alpha,\beta,\gamma,\theta,\psi,\varphi)} \right | \dd[6]{(\alpha,\beta,\gamma,\theta,\psi,\varphi)}.
\end{equation}
This expression can then be simplified by using the explicit components $D_{ij}=R\indices{_i^k}\lambda_{kl}R\indices{^l_j}$ in \eqref{eq:exp} and one finds
\be
N_{D}\rightarrow N_{\alpha,\beta,\gamma}=\frac{-3}{\sigma^2}  \left( \alpha^{2} + \beta^{2} + \gamma^{2} -\frac{1}{2}(\beta \gamma  + \alpha \beta +\alpha \gamma) \right) \label{ExpArg},
\ee
does not depend on the Euler angles of the rotation, and the Jacobian to transform \eqref{eq:pdf} is
\be
\left | \frac{\partial^6 D_{ij}}{\partial (\alpha,\beta,\gamma,\theta,\psi,\varphi)} \right | = (\alpha-\beta)(\alpha-\gamma)(\beta - \gamma) \sin \theta, 
\ee
which, surprisingly, only depends on one Euler angle $\theta$. 

After integration over the Euler angles the PDF becomes
\be
\begin{split}
\int\limits^{\pi}_0\int\limits^{\pi}_0\int\limits^{\pi}_0 &w(\alpha,\beta,\gamma,\theta,\psi,\varphi) \left | \frac{\partial^6 f_{ij}}{\partial (\alpha,\beta,\gamma,\theta,\psi,\varphi)} \right | \dd{\theta} \dd{\psi} \dd{\varphi} \dd[3]{(\alpha,\beta,\gamma)} \\
&=\frac{675 \sqrt{5}}{8\pi\sigma^6}\exp(N_{\alpha,\beta,\gamma})(\alpha-\beta)(\alpha-\gamma)(\beta - \gamma)\dd[3]{(\alpha,\beta,\gamma)}.\label{eq:Jacobian} 
% &= w(\alpha,\beta,\gamma)\dd^3(\alpha,\beta,\gamma).
\end{split}
\ee
It is important to know the range of validity of this PDF.  We want it to be positive semi-definite, so we notice that when $\gamma$ becomes greater than $\beta$, our PDF changes sign (and similarly for $\alpha$ and $\beta$).  Therefore (without loss of generality), we impose that 
$-\infty\leq \alpha\leq \beta\leq \gamma\leq \infty$, so that our PDF will be properly normalized\footnote{We note that this PDF essentially recovers the same result of \cite{Dorosh1970, Dorosh1978}, however our answer differs by a factor of five in the argument of the exponential, as Doroshkevich multiplied \eqref{eq:coma} by a factor of five to make the value of the first three diagonal components equal to the variance $\sigma^2$.}
\be \label{equalsone}
\int\limits^{\infty}_{-\infty} d\alpha \int\limits_{\alpha}^{\infty} d\beta \int\limits_{\beta}^{\infty} d\gamma \;  \left( \frac{675 \sqrt{5}}{8\pi\sigma^6} \right) \exp(N_{\alpha,\beta,\gamma})(\alpha-\beta)(\alpha-\gamma)(\beta - \gamma)=1.
\ee

Given this PDF we can now calculate the probability for spherical collapse, $\bcol$.
As discussed in Section \ref{Zapprox}, we need {\it at least} one negative eigenvalue for collapse to occur, and for spherical collapse we 
need all eigenvalues negative and approximately equal.  We can see the PDF is suppressed relative to a pure Gaussian distribution due to the
factors $(\alpha-\beta)(\alpha-\gamma)(\beta - \gamma)$ reflecting that it is rare to have perfectly spherical collapse, and forming objects like disks (pancakes) are more likely 
since typically one expects the eigenvalues to differ significantly. 
Recalling we have ordered the eigenvalues so that $\alpha$ is the most negative (and must be negative for collapse) we have for the probability of collapse to a PBH
\be
\bcol =  \int\limits^{0}_{-\infty} d\alpha \int\limits_{\alpha}^{\infty} d\beta \int\limits_{\beta}^{\infty} d\gamma \; \Theta\left[S(\alpha,\beta,\gamma)\right]  
W(\alpha,\beta,\gamma),
\ee
where $W$ is the integrand of \eqref{equalsone}, $\Theta$ is the Heaviside theta function, and $S$ is a ``shape function''.
As mentioned above, it is not enough that $\alpha<0$ for PBH formation, and the shape function places constraints on the values of $\beta$ and $\gamma$ for a given $\alpha$.
In \cite{KP81}, the authors supposed that a black hole could not form unless the matter clump was nearly spherical.  In this case, one obtains the following shape function
\be
S=\frac{\gamma+\beta}{2\alpha}-(1-x),
\ee
where $x\equiv r_g/r_1$ is the ratio of the Schwarzschild radius to the radius of the collapsing mass at the time it goes non-linear $\delta \sim {\cal O}(1)$.
One can determine that this limits both $\beta$ and $\gamma$ to be greater than $\alpha(1-x)$, yet also less than $\alpha$ (i.e. the initial deformation is nearly symmetric and the eigenvalues are approximately equal).  When $x \ll 1$ one can 
replace the argument of the exponent in the PDF with ${-9}\alpha^2/2$, and this recovers the result of\footnote{Performing the same integral with the full exponential argument leads to a complicated polynomial in $\arctan(x)$, but we have shown that this numeric result recovers that of  \cite{KP81} for $x\lesssim 0.2$.} \cite{KP81}, namely 
$\bcol = 0.02 \times x^5$.
However, other shape functions have been considered.  In particular, the authors of \cite{Harada2016} derived a different shape function.  There, instead of relying on a nearly spherical collapse, they made use of the ``hoop conjecture'' \cite{Malec2015,Misner1974}.  This conjecture states that a black hole will form from a distribution of mass if and only if the maximum circumference of the distribution is less than the Schwarzschild radius associated with its total mass.  Again using the $x \ll1$ approximation, an analytic estimate 
can be done and it was found $\bcol = 5.6 \times 10^{-2} \, x^5$ -- we will use this value in what follows. 

\subsection{Obtaining the Mass Fraction} \label{MassFrac}

To calculate the mass fraction of the universe contained within PBHs, we must consider two probabilities that determine whether an over-dense region will collapse and form a 
PBH.  The probability for sufficient spherical collapse was found above to be $\bcol = 5.6 \times 10^{-2} \, x^5$.  However, we also need to consider whether or not the {\it initial} inhomogeneity is conducive to collapse (which is also equivalent to determining whether caustics prevent PBH formation) -- we will call this probability $\bhom$.   
This probability describes the difference between the background and perturbed matter densities.  Using the Lemaitre-Tolman-Bondi (LTB) model \cite{LemTB,LTolB,LTBond}
(an exact solution of the Einstein equations) it is possible to capture this difference.  Using this one can characterize the formation of two important objects for the black hole: 1) the singularity, and 2) the apparent horizon.  The universe abhors naked singularities, so black hole formation can only occur when the apparent horizon forms before the singularity.   In the LTB formalism, we can define quantities that describe the degree of inhomogeneity for a region of space and a characteristic size of a dark matter clump in that region when the clump becomes non-linear in its density perturbations
\be
u=\frac{\rho_0-\overline{\rho}_1}{\overline{\rho}_1}, \quad x = \frac{r_g}{r_1},
\ee
where $\rho_0$ is the matter density at the center of the clump and $\overline{\rho}_1$ is the mean matter density inside $r_1$.  It was shown in \cite{KP81} that the formation of an apparent horizon precedes the formation of a singularity only if $u\lesssim x^{3/2}$.
Assuming that the inhomogeneity is a Gaussian and random variable one finds
\be
\label{oldBinhom}
\bhom \sim \frac{1}{\sigma}\int_0^{x^{3/2}}e^{-(u/\sigma)^2}\dd u\sim x^{3/2},
\ee
where it is again assumed that $x\ll1$.

This calculation has recently been refined in \cite{Kokubu2018}.  There it was argued that the result of \cite{KP81} slightly underestimates $\bhom$ due to the fact that the presence of a naked singularity can only reach the outside universe if a null signal propagating from the singularity can escape past an apparent horizon.  In other words, the time between singularity and apparent horizon formation must be less than the time it takes a null vector to traverse the distance between these objects.  An analytic result can be found 
in the limit of small initial perturbations
\be
\bhom \simeq 3.7 \, x^{3/2}, \quad x\ll1.
\ee

Given this result, the total mass fraction in PBHs is then the product of the two probabilities
\be
\beta = \bhom \bcol \simeq  \left( 3.7 \, x^{3/2} \right) \times \left( 5.6 \times 10^{-2} \, x^5 \right) \simeq  0.2 \, \delta_m^{13/2},
\label{AccMF}
\ee
where in the last step we have replaced $x$ by the mass density fluctuation $\delta_m$ at Hubble radius crossing -- which follows from scaling arguments of the clump and perturbations \cite{KP81}. 
This result is about an order of magnitude larger than that used in previous PBH studies (e.g. \cite{Georg:2017mqk,Georg:2016yxa}).  When performing calculations within the context of a specific model (in this paper the model will be Co-Decaying dark matter), we use this mass fraction estimate to obtain concrete results for the mass fraction as a function of black hole mass. 

\section{PBHs from Co-decay \label{cdpbh}}

\subsection{Minimum PBH Mass}
As discussed above, the EMDE begins when the energy density of the dark sector surpasses the SM radiation,
corresponding to temperatures $T_{\SM}<T_{eq}=\xi m$.
A very conservative estimate of the minimum PBH mass is then found by assuming that the mass contained in the horizon volume at this time collapses to 
form a PBH \cite{Georg:2016yxa}.  As discussed above, lacking special features in the primordial power spectrum, prior to this time radiation pressure 
will prevent the growth of smaller PBHs.  The minimal PBH mass is then 
given by the horizon as \cite{Georg:2016yxa} 
\begin{equation} 
M_{\mbox{\tiny{min}}}=\frac{3 m_p^2}{H_{eq}} \label{eq:min},
\end{equation}
where $H_{eq}$ is the Hubble parameter when the dark sector becomes equal to the SM radiation.  We emphasize that this is a very conservative estimate and most likely underestimates the mass of the PBH given that at this time half of the energy density 
is still in radiation and the pressure could have a substantial effect.  As we will see, establishing the minimal mass is important for avoiding the strongest constraints on PBHs, and so the more massive the first PBHs, the better for avoiding the stringent bounds near 
$M_{\mbox{\tiny{PBH}}} \approx 10^{15}$ g.  That is, increasing the mass would only lead to weaker constraints.

From the Friedmann equation at this time 
\begin{equation}
3H_{eq}^2m_p^2=\rho_{\DS} + \rho_{\SM} = 2 \rho_{\SM}=\frac{\pi^2}{15} g_* T_{\SM}^4,
\end{equation}
using $T_{\SM}=T_{eq}=\xi m$ 
we find the Hubble parameter
\begin{equation}
H_{eq}=\sqrt{\frac{\pi^2}{45m_p^2}} g_*^{1/2} \left( \xi m\right)^2.  \label{eq:hubble}
\end{equation}
Using this result \eqref{eq:min} leads to the minimum PBH mass
\begin{equation}
\label{MinFiducial}
M_{\mbox{\tiny{min}}}=\frac{9\sqrt{5}}{\pi g_*^{1/2}}   \left( \frac{m_p^3  }{ m_{\mbox{\tiny{DM}}}^2\xi^2} \right)
= 7.36 \times 10^{-5} M_\odot \left(\frac{106.75}{g_*}\right)^{1/2}\left(\frac{100 \; \mbox{GeV}}{m_{\mbox{\tiny{DM}}}}\right)^{2}\left(\frac{.1}{\xi}\right)^{2},
%= .26 M_\odot \left(\frac{114.38}{g_*}\right)^{1/2}\left(\frac{\mbox{GeV}}{m_{\mbox{\tiny{DM}}}}\right)^{2}\left(\frac{.1}{\xi}\right)^{2}.
\end{equation}
where $M_\odot = 2.0 \times 10^{33}$ g is the solar-mass.

\subsection{Maximum PBH Mass}
The maximum mass of PBHs formed during a matter-dominated phase is determined by those perturbations collapsing just before reheating. Since the pressure is negligible in a matter-dominated phase, we must account for sub-horizon growth -- the most massive PBHs need not correspond to the horizon scale (see \cite{Georg:2016yxa,Georg:2017mqk}). Therefore, the largest PBHs will form when $\delta_M(t_r)\sim \mathcal{O}(1)$ where
\begin{equation} \label{maxmass}
\delta_M(t_r)=\delta_M(t_H)\left(\frac{a(t_r)}{a(t_H)}\right)=\delta_C\left(\frac{M_{\mbox{\tiny{max}}}}{M_C}\right)^{\frac{1-n}{6}}\left(\frac{a(t_r)}{a(t_H)}\right)\sim \mathcal{O}(1),
\end{equation}
where we normalize the density perturbation and mass to the CMB scale, $\delta_{\mbox{\tiny{C}}} \simeq 3.8\times10^{-6}$ and $M_{\mbox{\tiny{C}}} = 10^{57} h^{-1} \; \si{g}$, $h$ is the Hubble parameter in units of 100 km/s $\si{Mpc}^{-1}$, $n$ is the spectral tilt of the primordial power spectrum, and the labels $r$ and $H$ denote reheating (decay of B particles) and horizon crossing, respectively.  Note that we have also used the fact that if the perturbation magnitudes are seeded by the primordial power spectrum, we can write $\delta_M \sim M^{(1-n)/6}$ \cite{Georg:2016yxa}.  Later, this identification will allow us to write the black hole formation probability as a function of mass.  Since we are in an EMDE ($a(t)\propto t^{2/3}$), we can furthermore substitute expressions for the scale factors at the various times.  The Hubble time at the time of decay is $H^{-1}=\Gamma_{\mbox{\tiny{B}}}^{-1}$, and this along with the relation between the horizon mass and scale factor at the time of horizon crossing of the modes resulting in the most massive PBH (which in an EMDE does {\it not} correspond to the horizon at reheating) implies that \eqref{maxmass} is given by 
\be
M_{\mbox{\tiny{max}}}=\alpha^{\frac{1}{n+3}} \left( \frac{M_{\mbox{\tiny{C}}}}{m_p}\right)^{\frac{n-1}{n+3}} \left( \frac{m_p}{\Gamma_{\mbox{\tiny{B}}}}\right)^{\frac{4}{n+3}} m_p,
\ee
where $\alpha \simeq 6.9 \times 10^{-23}$.  Considering different values of the tilt of the power spectrum we find
\bea
\label{MaxFiducial}
M_{\mbox{\tiny{max}}} &=& 5.0 \times 10^{-2} \; M_\odot  \left( \frac{10^{-22} \; \mbox{GeV}}{\Gamma_{\mbox{\tiny{B}}}}\right)^{\frac{4}{n+3}} \bigskip \;\; \; \; n=1.4 \nonumber \\
&=& 6.3  \; M_\odot   \left( \frac{10^{-22} \; \mbox{GeV}}{\Gamma_{\mbox{\tiny{B}}}}\right)^{\frac{4}{n+3}} \bigskip \bigskip \bigskip \; n=1.8 \nonumber \\
&=&  52.5 \; M_\odot   \left( \frac{10^{-22} \; \mbox{GeV}}{\Gamma_{\mbox{\tiny{B}}}}\right)^{\frac{4}{n+3}} \bigskip \bigskip \;\;\; \;\; \; n=2.0 \nonumber \\
\eea
where our choice of the fiducial decay rate corresponds to a reheat temperature of $T_r=12.6$ MeV.

At this point, it is important to note that, for certain ranges of $\Gamma_{\mbox{\tiny{B}}}$ and $n$ (given $m_{\mbox{\tiny{A}}}$ and $\xi$), one could find that $M_{min}>M_{max}$!  Via inspection of 
\eqref{MinFiducial} and \eqref{MaxFiducial}, one learns that this occurs at low values of $n$ and large values of $\Gamma_{\mbox{\tiny{B}}}$.  Smaller values of $n$ imply that matter perturbations require more time to become non-linear and collapse to a black hole.  The duration of the matter dominated phase decreases with increasing decay rate (for a given dark matter mass).  Since the perturbations require more time to grow and yet have less time to do so, there are combinations of the $n$ and $\Gamma_{\mbox{\tiny{B}}}$ such that, in order to become non-linear, a perturbation would have to enter the horizon and begin growing before the onset of the matter dominated phase.  This implies that \textit{no} black holes can form in a matter dominated phase characterized by these values of $n$ and $\Gamma_{\mbox{\tiny{B}}}$.  Therefore, we make sure to exclude these portions of parameter space: for example, the $f_{\mbox{\tiny{PBH}}}/f_{\mbox{\tiny{A}}}$ curves are cut-off when they encounter these parameters (see Fig. \ref{TotalF}).  

\subsection{PBH Abundance}
The mass fraction in PBHs during a EMDE phase is given by the probability that a PBH forms. As discussed in Section \ref{MassFrac}, this probability for a given PBH mass $M$ is given by
\begin{equation}
\beta(t_f)=\frac{\rho_{\mbox{\tiny PBH}}(M)}{\rho_{\mbox{\tiny tot}}} \simeq \bhom\beta_{col} \simeq  0.2 \, \delta_m^{13/2}.
\end{equation}
However, this only tells us the mass fraction at the time of their formation.  In order to compare to $\Omega_{DM}$, we must calculate the present day mass fraction.  The decays of the dark sector particles will induce a dilution of PBHs after they form.  Including this dilution factor, the mass fraction at the time of reheating is given by \cite{Georg:2017mqk}
\begin{equation}
\beta(t_r)=\left(\frac{5}{3}\right)^{3/4}(\Gamma_{\mbox{\tiny{B}}} t_f)^{1/2}\beta(M),
\end{equation}
where $t_f$ is the time at which the black hole forms.  To further expand this formula, we use the fact that, during a matter dominated era, we can identify 
\be
t_H = \frac{M}{6 \pi^4 m_P^2},
\ee
where $t_H$ is the time at which a given perturbation mode enters the horizon and $M$ is the mass contained within a sphere of radius $H^{-1}$.  Therefore, we can substitute $t_f = t_H \tau$, where $\tau=t_f/t_H>1$.  The expression we obtain for $\beta(t_r)$ represents a lower bound on the mass fraction of dark matter contained within black holes.

To determine the total portion of dark matter residing within black holes, we cannot simply integrate this mass fraction.  To find the total mass fraction, we must account for the enhancement of the black hole density during the radiation dominated era that follows reheating.  The black hole mass fraction will then remain approximately constant after matter-radiation equality to the present day.  To this end, we calculate the function $\psi(M)$:
\be
\psi(M) = \frac{a_{eq}}{a_r} \frac{\beta_r(M)}{M}.
\ee
Denoting the scale factor today as $a_0$ and using that $a_{eq}/a_0 = (z_{eq}+1)^{-1}$, $a_{r}/a_0 \simeq T_0/T_r$, $z_{eq}=3365$, and $T_0\simeq 230 \, \si{\mu eV}$, we can determine $\psi$ for a few fiducial values of the spectral tilt: 
\bea
\label{psiVals}
\psi(M) &=& 3 \times 10^{-21} M_{\odot}^{-1} \left( \frac{10.75}{g_\ast}\right)^{1/4} \left( \frac{\Gamma_{\mbox{\tiny{B}}}}{10^{-22} \si{GeV}}\right) \left(\frac{M_{\odot}}{M} \right)^{-\frac{13n_s-7}{12}} \quad \quad n=1.4 \nonumber \\
&=& 7 \times 10^{-11} M_{\odot}^{-1} \left( \frac{10.75}{g_\ast}\right)^{1/4} \left( \frac{\Gamma_{\mbox{\tiny{B}}}}{10^{-22} \si{GeV}}\right) \left(\frac{M_{\odot}}{M} \right)^{-\frac{13n_s-7}{12}} \quad \quad n=1.8 \nonumber \\
&=& 1 \times 10^{-5} M_{\odot}^{-1} \left( \frac{10.75}{g_\ast}\right)^{1/4} \left( \frac{\Gamma_{\mbox{\tiny{B}}}}{10^{-22} \si{GeV}}\right) \left(\frac{M_{\odot}}{M} \right)^{-\frac{13n_s-7}{12}} \quad \quad n=2 \nonumber \\
\eea

\begin{figure}[t!]
\begin{center}
\includegraphics[height=4in]{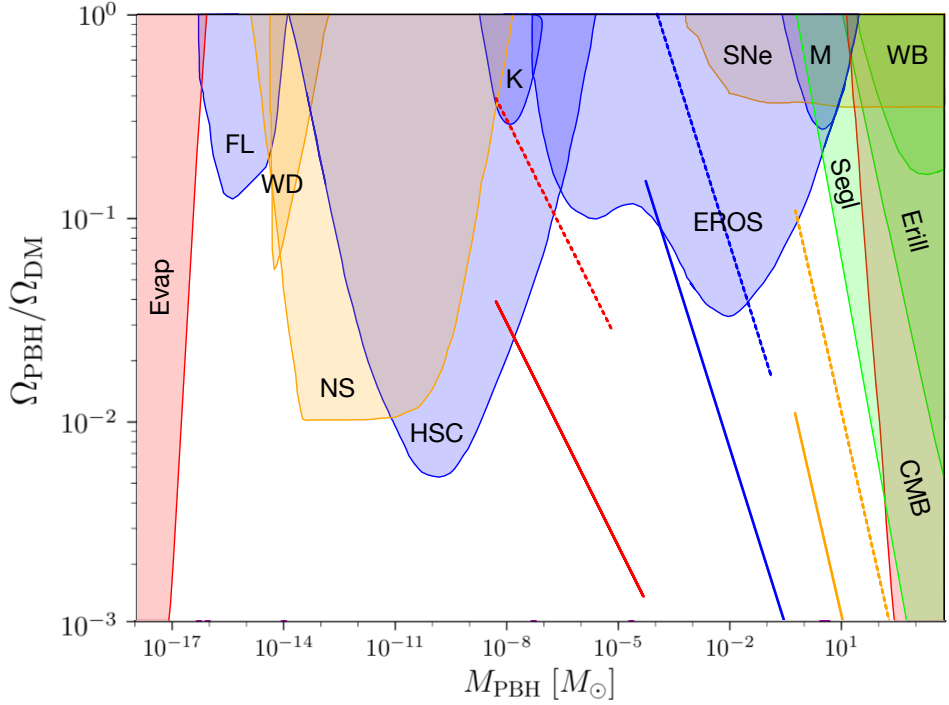}
\end{center}
\caption{Constraints on the abundance of PBHs in the Co-Decay model.  We plot the mass distribution function, $\psi$, against the monochromatic black hole constraints.  The red solid line corresponds to $n_s=1.8$, $m_A=10^4$ GeV, and $\Gamma_{\mbox{\tiny{B}}}=10^{-16}$ GeV. The blue solid line corresponds to $n_s=2$, $m_A=10^2$ GeV, and $\Gamma_{\mbox{\tiny{B}}}=10^{-20}$ GeV. The orange solid line corresponds to $n_s=2.2$, $m_A=1$ GeV, and $\Gamma_{\mbox{\tiny{B}}}=10^{-24}$ GeV.  The dashed lines correspond to decay rates 10 times the value of their solid counterparts. We use the \textbf{\={A}BC} data set from \cite{Lehmann2018}. The labeled constraints are from BH evaporation (\texttt{evap}, \cite{Carr:2009jm}), GRB femtolensing observations (\texttt{FL}, \cite{Barnacka:2012bm}), white dwarf explosions (\texttt{WD}, \cite{Graham:2015apa}), Hyper Suprime-Cam (\texttt{HSC}, \cite{Niikura:2017zjd}), Kepler (\texttt{K}, \cite{Griest:2013aaa}), EROS-II (\texttt{EROS}, \cite{Tisserand:2006zx}), supernova lensing (\texttt{SNe}, \cite{Zumalacarregui:2017qqd}), MACHO (\texttt{MACHO}, \cite{Allsman:2000kg}), Segue I dynamics (\texttt{SegI}, \cite{Koushiappas:2017chw}), Eridanus II dynamics (\texttt{EriII}, \cite{Brandt:2016aco}), wide binary dynamics (\texttt{WB}, \cite{Monroy-Rodriguez:2014ula}), and CMB observables (\texttt{CMB}, \cite{Ali-Haimoud:2016mbv,Carr:2017jsz}).}
\label{fig4}
\end{figure}

In the past, all PBH constraints were cast in terms of ``monochromatic'' mass functions (see, for instance, \cite{Carr2016} for a discussion of these constraints).  This means that most constraints cited in various papers only apply when \textit{all} black holes are characterized by the same mass.  More general cosmological scenarios can lead to mass distribution functions that are extended in mass, like $\psi$ above.  Therefore, some recent work has been devoted to adapting the monochromatic constraints to extended mass functions \cite{Carr:2017jsz,Bellomo:2017zsr}.  This usually means that the constraints become more restrictive.  However, a recent work \cite{Lehmann2018} has shown that a maximum allowed PBH mass fraction can be derived \textit{independently} of a given mass distribution shape.  When we restrict $\psi(M)$ for various parameters of the Co-Decay model, we can thus determine the models viability in two ways: (1) find whether $\psi(M)$ falls into an area excluded by monochromatic constraints, and (2) determining whether the total mass fraction (see below) is larger than that allowed by analysis similar to \cite{Lehmann2018}.  If either of these scenarios occur, Co-Decay with those parameters is excluded (see Fig. \ref{fig4} for comparisons between the constraints and $\psi$ for various parameters).

\begin{figure}
\begin{center}
\includegraphics{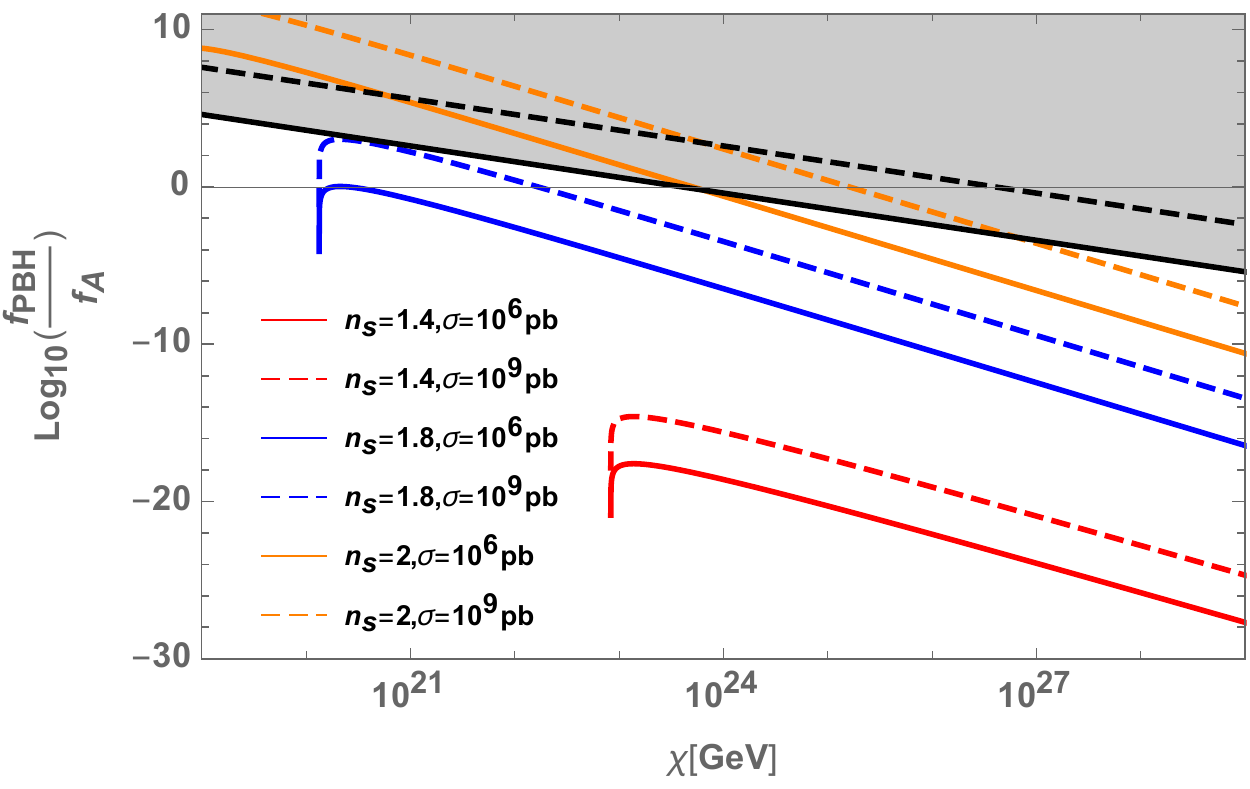}
\end{center}
\caption{The ratio of the total abundances of PBHs to A particle dark matter shown for various spectral tilts, threshold annihilation rates, and a dark matter mass of 100 GeV.  The black lines correspond to maximum allowed mass fractions of primordial black holes.  For spectral tilts of 1.4 and 1.8, there is a value of $\chi$ for which no black hole formation occurs and $f_{\mbox{\tiny{PBH}}}$ rapidly approaches zero at this value.}
\label{TotalF}
\end{figure}

The quantity $\psi(M)dM$ represents the PBH dark matter fraction in the interval $(M,M+dM)$, which means that \cite{Carr2017}
\be
f_{\mbox{\tiny{PBH}}}\equiv\frac{\Omega_{\mbox{\tiny{PBH}}}}{\Omega_{\mbox{\tiny{DM}}}} = \int^{M_{max}}_{M_{max}} dM \psi(M).
\ee
Upon integration, we find that we can restrict the total fraction in terms of the $\chi$ parameter defined above (see Fig. \ref{TotalF}).  However, in this case, we must still pick a value for $m_{\mbox{\tiny{A}}}$, which in turn implies that $\chi$ here plays the role of $\Gamma_{\mbox{\tiny{B}}}$.   

Our constraints on the Co-Decay PBHs appear in Figure \ref{TotalF}.
We see from Fig.~\ref{TotalF} that as $n_s$ increases, so does the portion of dark matter residing in black holes.  This makes sense because the larger the spectral tilt, the larger the density contrast a given mass scale has upon horizon entry.  The difference between the solid ($\sigma = 10^6$ pb) and dashed lines ($\sigma = 10^9$ pb) resides in the fact that for a given mass and decay rate a higher annihilation threshold implies a smaller A particle abundance (see also Fig. \ref{Abund}).  The black solid and dashed lines represent the maximum ratio as derived from \cite{Lehmann2018}.  We use the most restrictive value for the maximum allowed total, which is determined from the constraint set \textbf{\={A}BC} from \cite{Lehmann2018}.  

There are two interesting features in the ratio dependence of $\chi$.  The most obvious of these is the sharp downturn to zero at certain values of low $\chi$ (large $\Gamma_{\mbox{\tiny{B}}}$).  For each value of $n_s$, this occurs at the value of the decay rate for which the PBH maximum mass approaches the minimum mass.  As discussed above, this implies that there is no black hole formation throughout the matter dominated era.  The more surprising of the features is the gradual decrease in mass fraction for increasing $\chi$ (decreasing $\Gamma_{\mbox{\tiny{B}}}$).  Decreasing decay rate implies both a lengthening of the matter domination and an increase in the maximum mass for the black holes.  This suggests that for smaller decay rates, there would be more time to form black holes and a larger range of masses that can be collapsed to black holes.  

Two factors contribute to the decreasing nature of the total mass fraction.  As can be seen from \eqref{aAbund}, the A particle abundance is inversely proportional to the decay rate.  As the EMDE lengthens, the amount of relic A particles increase as well, which would cause a decrease in the ratio of the mass fractions.  Secondly, the mass distribution function is determined by the probability of collapse at the time of reheating.  Due to the microphysics of Co-Decay, the black holes formed early in the matter dominated phase are diluted due to the entropy transfers between the dark and visible sectors.  For this reason, the longer the matter domination, the more dilution of the relics that occurs.

We would now like to consider whether there are portions of the allowed parameter space in which PBHs constitute an appreciable fraction of the universe's DM budget.  To this end, we recall from Fig. \ref{Abund} that $f_A=1$ at $\chi=10^{24}$ GeV for $\sigma=10^6$ pb, and at $\chi=10^{27}$ GeV for $\sigma=10^9$ pb.  For $n_s<2$, PBHs never make up more than a negligible portion of DM. We can see this from the fact that for the larger cross-section values, the maximum fraction of DM in black holes is $10^4f_A$, and occurs at $\chi\sim10^{21}$.  The value of $f_A$ at this point is $\sim 10^{-8}$.  At a spectral tilt of 1.8, Co-Decay can only account for \textit{one ten-thousandth} of all DM.

When we push the spectral tilt up to 2 and have a dark sector annihilation rate of $10^9$ pb, we see that if $\chi\in(10^{24},10^{27})$ GeV, PBHs can account for all of DM, while still avoiding the constraints in Fig. \ref{fig4}: $f_{\mbox{\tiny{PBH}}}=10^4 f_A = 1$ at $\chi=10^{24}$ GeV.  When $\chi\sim\mathcal{O}(10^{25})$ GeV, the A particles and PBHs have equal abundances, but their total is only $\mathcal{O}(.01)$.  Moving to higher $\chi$ (lower $\Gamma_{\mbox{\tiny{B}}}$) brings us closer to the point where $f_A=1$.  When that is the case, Fig. \ref{TotalF} still shows that PBHs can account for $\sim\mathcal{O}(.01)-\mathcal{O}(.1)$ of DM.  

The above plot and reasoning leads us to suspect that, although it is unlikely that PBHs are all of DM, early-forming, solar-mass black holes can be non-negligible fractions of the total DM budget for appreciable ranges of parameter space.  Thus, Co-Decay DM offers an interesting explanation for the presence of an abundance of near solar-mass black holes.  

\section{Conclusions}
In this paper, we have considered the production of PBHs in models of Co-Decay. We found the mass fraction and relic abundance for varying values of the primordial power spectrum. We have seen that observational constraints place important restrictions on the parameter space for Co-Decay and the amount of PBHs that can comprise DM. We find that given the range of masses that are produced that the mass range near or around a solar-mass is the most promising. Depending on the further evolution of these PBHs, this could provide a new mechanism for accounting for near solar-mass black holes, like those detected by LIGO.  An important challenge is then to see if these PBHs can evolve to form binary pairs and determine their merger rate. We leave this to future work.

\section*{Acknowledgements}
We thank Jeff Dror, Bhaskar Dutta, Adrienne Erickcek, Benjamin Lehmann, Eric Kufick, Alex Kusenko, and Stefano Profumo for useful discussions. This research was supported in part by NASA Astrophysics Theory Grant NNH12ZDA001N and DOE grant DE-FG02-85ER40237.

\bibliographystyle{apsrev4-1}

\begin{thebibliography}{73}%
\makeatletter
\providecommand \@ifxundefined [1]{%
 \@ifx{#1\undefined}
}%
\providecommand \@ifnum [1]{%
 \ifnum #1\expandafter \@firstoftwo
 \else \expandafter \@secondoftwo
 \fi
}%
\providecommand \@ifx [1]{%
 \ifx #1\expandafter \@firstoftwo
 \else \expandafter \@secondoftwo
 \fi
}%
\providecommand \natexlab [1]{#1}%
\providecommand \enquote  [1]{``#1''}%
\providecommand \bibnamefont  [1]{#1}%
\providecommand \bibfnamefont [1]{#1}%
\providecommand \citenamefont [1]{#1}%
\providecommand \href@noop [0]{\@secondoftwo}%
\providecommand \href [0]{\begingroup \@sanitize@url \@href}%
\providecommand \@href[1]{\@@startlink{#1}\@@href}%
\providecommand \@@href[1]{\endgroup#1\@@endlink}%
\providecommand \@sanitize@url [0]{\catcode `\\12\catcode `\$12\catcode
  `\&12\catcode `\#12\catcode `\^12\catcode `\_12\catcode `\%12\relax}%
\providecommand \@@startlink[1]{}%
\providecommand \@@endlink[0]{}%
\providecommand \url  [0]{\begingroup\@sanitize@url \@url }%
\providecommand \@url [1]{\endgroup\@href {#1}{\urlprefix }}%
\providecommand \urlprefix  [0]{URL }%
\providecommand \Eprint [0]{\href }%
\providecommand \doibase [0]{http://dx.doi.org/}%
\providecommand \selectlanguage [0]{\@gobble}%
\providecommand \bibinfo  [0]{\@secondoftwo}%
\providecommand \bibfield  [0]{\@secondoftwo}%
\providecommand \translation [1]{[#1]}%
\providecommand \BibitemOpen [0]{}%
\providecommand \bibitemStop [0]{}%
\providecommand \bibitemNoStop [0]{.\EOS\space}%
\providecommand \EOS [0]{\spacefactor3000\relax}%
\providecommand \BibitemShut  [1]{\csname bibitem#1\endcsname}%
\let\auto@bib@innerbib\@empty
%</preamble>
\bibitem [{\citenamefont {Tanabashi}\ \emph {et~al.}(2018)\citenamefont
  {Tanabashi} \emph {et~al.}}]{Tanabashi:2018oca}%
  \BibitemOpen
  \bibfield  {author} {\bibinfo {author} {\bibfnamefont {M.}~\bibnamefont
  {Tanabashi}} \emph {et~al.} (\bibinfo {collaboration} {Particle Data
  Group}),\ }\href {\doibase 10.1103/PhysRevD.98.030001} {\bibfield  {journal}
  {\bibinfo  {journal} {Phys. Rev.}\ }\textbf {\bibinfo {volume} {D98}},\
  \bibinfo {pages} {030001} (\bibinfo {year} {2018})}\BibitemShut {NoStop}%
%%CITATION = PHRVA,D98,030001;%%
\bibitem [{\citenamefont {Kane}\ \emph {et~al.}(2015)\citenamefont {Kane},
  \citenamefont {Sinha},\ and\ \citenamefont {Watson}}]{Kane:2015jia}%
  \BibitemOpen
  \bibfield  {author} {\bibinfo {author} {\bibfnamefont {G.}~\bibnamefont
  {Kane}}, \bibinfo {author} {\bibfnamefont {K.}~\bibnamefont {Sinha}}, \ and\
  \bibinfo {author} {\bibfnamefont {S.}~\bibnamefont {Watson}},\ }\href
  {\doibase 10.1142/S0218271815300220} {\bibfield  {journal} {\bibinfo
  {journal} {Int. J. Mod. Phys.}\ }\textbf {\bibinfo {volume} {D24}},\ \bibinfo
  {pages} {1530022} (\bibinfo {year} {2015})},\ \Eprint
  {http://arxiv.org/abs/1502.07746} {arXiv:1502.07746 [hep-th]} \BibitemShut
  {NoStop}%
%%CITATION = ARXIV:1502.07746;%%
\bibitem [{\citenamefont {Adshead}\ \emph {et~al.}(2016)\citenamefont
  {Adshead}, \citenamefont {Cui},\ and\ \citenamefont
  {Shelton}}]{Adshead:2016xxj}%
  \BibitemOpen
  \bibfield  {author} {\bibinfo {author} {\bibfnamefont {P.}~\bibnamefont
  {Adshead}}, \bibinfo {author} {\bibfnamefont {Y.}~\bibnamefont {Cui}}, \ and\
  \bibinfo {author} {\bibfnamefont {J.}~\bibnamefont {Shelton}},\ }\href
  {\doibase 10.1007/JHEP06(2016)016} {\bibfield  {journal} {\bibinfo  {journal}
  {JHEP}\ }\textbf {\bibinfo {volume} {06}},\ \bibinfo {pages} {016} (\bibinfo
  {year} {2016})},\ \Eprint {http://arxiv.org/abs/1604.02458} {arXiv:1604.02458
  [hep-ph]} \BibitemShut {NoStop}%
%%CITATION = ARXIV:1604.02458;%%
\bibitem [{\citenamefont {Abazajian}\ \emph {et~al.}(2016)\citenamefont
  {Abazajian} \emph {et~al.}}]{Abazajian:2016yjj}%
  \BibitemOpen
  \bibfield  {author} {\bibinfo {author} {\bibfnamefont {K.~N.}\ \bibnamefont
  {Abazajian}} \emph {et~al.} (\bibinfo {collaboration} {CMB-S4}),\ }\href@noop
  {} {\  (\bibinfo {year} {2016})},\ \Eprint {http://arxiv.org/abs/1610.02743}
  {arXiv:1610.02743 [astro-ph.CO]} \BibitemShut {NoStop}%
%%CITATION = ARXIV:1610.02743;%%
\bibitem [{\citenamefont {Dror}\ \emph {et~al.}(2016)\citenamefont {Dror},
  \citenamefont {Kuflik},\ and\ \citenamefont {Ng}}]{Dror:2016rxc}%
  \BibitemOpen
  \bibfield  {author} {\bibinfo {author} {\bibfnamefont {J.~A.}\ \bibnamefont
  {Dror}}, \bibinfo {author} {\bibfnamefont {E.}~\bibnamefont {Kuflik}}, \ and\
  \bibinfo {author} {\bibfnamefont {W.~H.}\ \bibnamefont {Ng}},\ }\href
  {\doibase 10.1103/PhysRevLett.117.211801} {\bibfield  {journal} {\bibinfo
  {journal} {Phys. Rev. Lett.}\ }\textbf {\bibinfo {volume} {117}},\ \bibinfo
  {pages} {211801} (\bibinfo {year} {2016})},\ \Eprint
  {http://arxiv.org/abs/1607.03110} {arXiv:1607.03110 [hep-ph]} \BibitemShut
  {NoStop}%
%%CITATION = ARXIV:1607.03110;%%
\bibitem [{\citenamefont {Dror}\ \emph {et~al.}(2018)\citenamefont {Dror},
  \citenamefont {Kuflik}, \citenamefont {Melcher},\ and\ \citenamefont
  {Watson}}]{Dror:2017gjq}%
  \BibitemOpen
  \bibfield  {author} {\bibinfo {author} {\bibfnamefont {J.~A.}\ \bibnamefont
  {Dror}}, \bibinfo {author} {\bibfnamefont {E.}~\bibnamefont {Kuflik}},
  \bibinfo {author} {\bibfnamefont {B.}~\bibnamefont {Melcher}}, \ and\
  \bibinfo {author} {\bibfnamefont {S.}~\bibnamefont {Watson}},\ }\href
  {\doibase 10.1103/PhysRevD.97.063524} {\bibfield  {journal} {\bibinfo
  {journal} {Phys. Rev.}\ }\textbf {\bibinfo {volume} {D97}},\ \bibinfo {pages}
  {063524} (\bibinfo {year} {2018})},\ \Eprint
  {http://arxiv.org/abs/1711.04773} {arXiv:1711.04773 [hep-ph]} \BibitemShut
  {NoStop}%
%%CITATION = ARXIV:1711.04773;%%
\bibitem [{\citenamefont {Dery}\ \emph {et~al.}(2019)\citenamefont {Dery},
  \citenamefont {Dror}, \citenamefont {Stephenson~Haskins}, \citenamefont
  {Hochberg},\ and\ \citenamefont {Kuflik}}]{Dery:2019jwf}%
  \BibitemOpen
  \bibfield  {author} {\bibinfo {author} {\bibfnamefont {A.}~\bibnamefont
  {Dery}}, \bibinfo {author} {\bibfnamefont {J.~A.}\ \bibnamefont {Dror}},
  \bibinfo {author} {\bibfnamefont {L.}~\bibnamefont {Stephenson~Haskins}},
  \bibinfo {author} {\bibfnamefont {Y.}~\bibnamefont {Hochberg}}, \ and\
  \bibinfo {author} {\bibfnamefont {E.}~\bibnamefont {Kuflik}},\ }\href@noop {}
  {\  (\bibinfo {year} {2019})},\ \Eprint {http://arxiv.org/abs/1901.02018}
  {arXiv:1901.02018 [hep-ph]} \BibitemShut {NoStop}%
%%CITATION = ARXIV:1901.02018;%%
\bibitem [{\citenamefont {Pappadopulo}\ \emph {et~al.}(2016)\citenamefont
  {Pappadopulo}, \citenamefont {Ruderman},\ and\ \citenamefont
  {Trevisan}}]{Pappadopulo:2016pkp}%
  \BibitemOpen
  \bibfield  {author} {\bibinfo {author} {\bibfnamefont {D.}~\bibnamefont
  {Pappadopulo}}, \bibinfo {author} {\bibfnamefont {J.~T.}\ \bibnamefont
  {Ruderman}}, \ and\ \bibinfo {author} {\bibfnamefont {G.}~\bibnamefont
  {Trevisan}},\ }\href {\doibase 10.1103/PhysRevD.94.035005} {\bibfield
  {journal} {\bibinfo  {journal} {Phys. Rev.}\ }\textbf {\bibinfo {volume}
  {D94}},\ \bibinfo {pages} {035005} (\bibinfo {year} {2016})},\ \Eprint
  {http://arxiv.org/abs/1602.04219} {arXiv:1602.04219 [hep-ph]} \BibitemShut
  {NoStop}%
%%CITATION = ARXIV:1602.04219;%%
\bibitem [{\citenamefont {Farina}\ \emph {et~al.}(2016)\citenamefont {Farina},
  \citenamefont {Pappadopulo}, \citenamefont {Ruderman},\ and\ \citenamefont
  {Trevisan}}]{Farina:2016llk}%
  \BibitemOpen
  \bibfield  {author} {\bibinfo {author} {\bibfnamefont {M.}~\bibnamefont
  {Farina}}, \bibinfo {author} {\bibfnamefont {D.}~\bibnamefont {Pappadopulo}},
  \bibinfo {author} {\bibfnamefont {J.~T.}\ \bibnamefont {Ruderman}}, \ and\
  \bibinfo {author} {\bibfnamefont {G.}~\bibnamefont {Trevisan}},\ }\href
  {\doibase 10.1007/JHEP12(2016)039} {\bibfield  {journal} {\bibinfo  {journal}
  {JHEP}\ }\textbf {\bibinfo {volume} {12}},\ \bibinfo {pages} {039} (\bibinfo
  {year} {2016})},\ \Eprint {http://arxiv.org/abs/1607.03108} {arXiv:1607.03108
  [hep-ph]} \BibitemShut {NoStop}%
%%CITATION = ARXIV:1607.03108;%%
\bibitem [{\citenamefont {Erickcek}\ and\ \citenamefont
  {Sigurdson}(2011)}]{Erickcek:2011us}%
  \BibitemOpen
  \bibfield  {author} {\bibinfo {author} {\bibfnamefont {A.~L.}\ \bibnamefont
  {Erickcek}}\ and\ \bibinfo {author} {\bibfnamefont {K.}~\bibnamefont
  {Sigurdson}},\ }\href {\doibase 10.1103/PhysRevD.84.083503} {\bibfield
  {journal} {\bibinfo  {journal} {Phys. Rev.}\ }\textbf {\bibinfo {volume}
  {D84}},\ \bibinfo {pages} {083503} (\bibinfo {year} {2011})},\ \Eprint
  {http://arxiv.org/abs/1106.0536} {arXiv:1106.0536 [astro-ph.CO]} \BibitemShut
  {NoStop}%
%%CITATION = ARXIV:1106.0536;%%
\bibitem [{\citenamefont {Fan}\ \emph {et~al.}(2014)\citenamefont {Fan},
  \citenamefont {Ozsoy},\ and\ \citenamefont {Watson}}]{Fan:2014zua}%
  \BibitemOpen
  \bibfield  {author} {\bibinfo {author} {\bibfnamefont {J.}~\bibnamefont
  {Fan}}, \bibinfo {author} {\bibfnamefont {O.}~\bibnamefont {Ozsoy}}, \ and\
  \bibinfo {author} {\bibfnamefont {S.}~\bibnamefont {Watson}},\ }\href
  {\doibase 10.1103/PhysRevD.90.043536} {\bibfield  {journal} {\bibinfo
  {journal} {Phys. Rev.}\ }\textbf {\bibinfo {volume} {D90}},\ \bibinfo {pages}
  {043536} (\bibinfo {year} {2014})},\ \Eprint {http://arxiv.org/abs/1405.7373}
  {arXiv:1405.7373 [hep-ph]} \BibitemShut {NoStop}%
%%CITATION = ARXIV:1405.7373;%%
\bibitem [{\citenamefont {Erickcek}\ \emph {et~al.}(2016)\citenamefont
  {Erickcek}, \citenamefont {Sinha},\ and\ \citenamefont
  {Watson}}]{Erickcek:2015bda}%
  \BibitemOpen
  \bibfield  {author} {\bibinfo {author} {\bibfnamefont {A.~L.}\ \bibnamefont
  {Erickcek}}, \bibinfo {author} {\bibfnamefont {K.}~\bibnamefont {Sinha}}, \
  and\ \bibinfo {author} {\bibfnamefont {S.}~\bibnamefont {Watson}},\ }\href
  {\doibase 10.1103/PhysRevD.94.063502} {\bibfield  {journal} {\bibinfo
  {journal} {Phys. Rev.}\ }\textbf {\bibinfo {volume} {D94}},\ \bibinfo {pages}
  {063502} (\bibinfo {year} {2016})},\ \Eprint
  {http://arxiv.org/abs/1510.04291} {arXiv:1510.04291 [hep-ph]} \BibitemShut
  {NoStop}%
%%CITATION = ARXIV:1510.04291;%%
\bibitem [{\citenamefont {Georg}\ \emph {et~al.}(2016)\citenamefont {Georg},
  \citenamefont {Sengor},\ and\ \citenamefont {Watson}}]{Georg:2016yxa}%
  \BibitemOpen
  \bibfield  {author} {\bibinfo {author} {\bibfnamefont {J.}~\bibnamefont
  {Georg}}, \bibinfo {author} {\bibfnamefont {G.}~\bibnamefont {Sengor}}, \
  and\ \bibinfo {author} {\bibfnamefont {S.}~\bibnamefont {Watson}},\ }\href
  {\doibase 10.1103/PhysRevD.93.123523} {\bibfield  {journal} {\bibinfo
  {journal} {Phys. Rev.}\ }\textbf {\bibinfo {volume} {D93}},\ \bibinfo {pages}
  {123523} (\bibinfo {year} {2016})},\ \Eprint
  {http://arxiv.org/abs/1603.00023} {arXiv:1603.00023 [hep-ph]} \BibitemShut
  {NoStop}%
%%CITATION = ARXIV:1603.00023;%%
\bibitem [{\citenamefont {Georg}\ and\ \citenamefont
  {Watson}(2017)}]{Georg:2017mqk}%
  \BibitemOpen
  \bibfield  {author} {\bibinfo {author} {\bibfnamefont {J.}~\bibnamefont
  {Georg}}\ and\ \bibinfo {author} {\bibfnamefont {S.}~\bibnamefont {Watson}},\
  }\href {\doibase 10.1007/JHEP09(2017)138} {\bibfield  {journal} {\bibinfo
  {journal} {JHEP}\ }\textbf {\bibinfo {volume} {09}},\ \bibinfo {pages} {138}
  (\bibinfo {year} {2017})},\ \bibinfo {note} {[JHEP09,138(2017)]},\ \Eprint
  {http://arxiv.org/abs/1703.04825} {arXiv:1703.04825 [astro-ph.CO]}
  \BibitemShut {NoStop}%
%%CITATION = ARXIV:1703.04825;%%
\bibitem [{\citenamefont {Cotner}\ and\ \citenamefont
  {Kusenko}(2017{\natexlab{a}})}]{Cotner:2017tir}%
  \BibitemOpen
  \bibfield  {author} {\bibinfo {author} {\bibfnamefont {E.}~\bibnamefont
  {Cotner}}\ and\ \bibinfo {author} {\bibfnamefont {A.}~\bibnamefont
  {Kusenko}},\ }\href {\doibase 10.1103/PhysRevD.96.103002} {\bibfield
  {journal} {\bibinfo  {journal} {Phys. Rev.}\ }\textbf {\bibinfo {volume}
  {D96}},\ \bibinfo {pages} {103002} (\bibinfo {year} {2017}{\natexlab{a}})},\
  \Eprint {http://arxiv.org/abs/1706.09003} {arXiv:1706.09003 [astro-ph.CO]}
  \BibitemShut {NoStop}%
%%CITATION = ARXIV:1706.09003;%%
\bibitem [{\citenamefont {Cotner}\ and\ \citenamefont
  {Kusenko}(2017{\natexlab{b}})}]{Cotner:2016cvr}%
  \BibitemOpen
  \bibfield  {author} {\bibinfo {author} {\bibfnamefont {E.}~\bibnamefont
  {Cotner}}\ and\ \bibinfo {author} {\bibfnamefont {A.}~\bibnamefont
  {Kusenko}},\ }\href {\doibase 10.1103/PhysRevLett.119.031103} {\bibfield
  {journal} {\bibinfo  {journal} {Phys. Rev. Lett.}\ }\textbf {\bibinfo
  {volume} {119}},\ \bibinfo {pages} {031103} (\bibinfo {year}
  {2017}{\natexlab{b}})},\ \Eprint {http://arxiv.org/abs/1612.02529}
  {arXiv:1612.02529 [astro-ph.CO]} \BibitemShut {NoStop}%
%%CITATION = ARXIV:1612.02529;%%
\bibitem [{\citenamefont {Allahverdi}\ \emph {et~al.}(2018)\citenamefont
  {Allahverdi}, \citenamefont {Dent},\ and\ \citenamefont
  {Osinski}}]{Allahverdi:2017sks}%
  \BibitemOpen
  \bibfield  {author} {\bibinfo {author} {\bibfnamefont {R.}~\bibnamefont
  {Allahverdi}}, \bibinfo {author} {\bibfnamefont {J.}~\bibnamefont {Dent}}, \
  and\ \bibinfo {author} {\bibfnamefont {J.}~\bibnamefont {Osinski}},\ }\href
  {\doibase 10.1103/PhysRevD.97.055013} {\bibfield  {journal} {\bibinfo
  {journal} {Phys. Rev.}\ }\textbf {\bibinfo {volume} {D97}},\ \bibinfo {pages}
  {055013} (\bibinfo {year} {2018})},\ \Eprint
  {http://arxiv.org/abs/1711.10511} {arXiv:1711.10511 [astro-ph.CO]}
  \BibitemShut {NoStop}%
%%CITATION = ARXIV:1711.10511;%%
\bibitem [{\citenamefont {Byrnes}\ \emph {et~al.}(2018)\citenamefont {Byrnes},
  \citenamefont {Cole},\ and\ \citenamefont {Patil}}]{Byrnes:2018txb}%
  \BibitemOpen
  \bibfield  {author} {\bibinfo {author} {\bibfnamefont {C.~T.}\ \bibnamefont
  {Byrnes}}, \bibinfo {author} {\bibfnamefont {P.~S.}\ \bibnamefont {Cole}}, \
  and\ \bibinfo {author} {\bibfnamefont {S.~P.}\ \bibnamefont {Patil}},\
  }\href@noop {} {\  (\bibinfo {year} {2018})},\ \Eprint
  {http://arxiv.org/abs/1811.11158} {arXiv:1811.11158 [astro-ph.CO]}
  \BibitemShut {NoStop}%
%%CITATION = ARXIV:1811.11158;%%
\bibitem [{\citenamefont {Kawasaki}\ and\ \citenamefont
  {Takhistov}(2018)}]{Kawasaki:2018daf}%
  \BibitemOpen
  \bibfield  {author} {\bibinfo {author} {\bibfnamefont {M.}~\bibnamefont
  {Kawasaki}}\ and\ \bibinfo {author} {\bibfnamefont {V.}~\bibnamefont
  {Takhistov}},\ }\href {\doibase 10.1103/PhysRevD.98.123514} {\bibfield
  {journal} {\bibinfo  {journal} {Phys. Rev.}\ }\textbf {\bibinfo {volume}
  {D98}},\ \bibinfo {pages} {123514} (\bibinfo {year} {2018})},\ \Eprint
  {http://arxiv.org/abs/1810.02547} {arXiv:1810.02547 [hep-th]} \BibitemShut
  {NoStop}%
%%CITATION = ARXIV:1810.02547;%%
\bibitem [{\citenamefont {Cai}\ \emph {et~al.}(2018)\citenamefont {Cai},
  \citenamefont {Tong}, \citenamefont {Wang},\ and\ \citenamefont
  {Yan}}]{Cai:2018tuh}%
  \BibitemOpen
  \bibfield  {author} {\bibinfo {author} {\bibfnamefont {Y.-F.}\ \bibnamefont
  {Cai}}, \bibinfo {author} {\bibfnamefont {X.}~\bibnamefont {Tong}}, \bibinfo
  {author} {\bibfnamefont {D.-G.}\ \bibnamefont {Wang}}, \ and\ \bibinfo
  {author} {\bibfnamefont {S.-F.}\ \bibnamefont {Yan}},\ }\href {\doibase
  10.1103/PhysRevLett.121.081306} {\bibfield  {journal} {\bibinfo  {journal}
  {Phys. Rev. Lett.}\ }\textbf {\bibinfo {volume} {121}},\ \bibinfo {pages}
  {081306} (\bibinfo {year} {2018})},\ \Eprint
  {http://arxiv.org/abs/1805.03639} {arXiv:1805.03639 [astro-ph.CO]}
  \BibitemShut {NoStop}%
%%CITATION = ARXIV:1805.03639;%%
\bibitem [{\citenamefont {Kohri}\ and\ \citenamefont
  {Terada}(2018)}]{Kohri:2018qtx}%
  \BibitemOpen
  \bibfield  {author} {\bibinfo {author} {\bibfnamefont {K.}~\bibnamefont
  {Kohri}}\ and\ \bibinfo {author} {\bibfnamefont {T.}~\bibnamefont {Terada}},\
  }\href {\doibase 10.1088/1361-6382/aaea18} {\bibfield  {journal} {\bibinfo
  {journal} {Class. Quant. Grav.}\ }\textbf {\bibinfo {volume} {35}},\ \bibinfo
  {pages} {235017} (\bibinfo {year} {2018})},\ \Eprint
  {http://arxiv.org/abs/1802.06785} {arXiv:1802.06785 [astro-ph.CO]}
  \BibitemShut {NoStop}%
%%CITATION = ARXIV:1802.06785;%%
\bibitem [{\citenamefont {Cotner}\ \emph {et~al.}(2018)\citenamefont {Cotner},
  \citenamefont {Kusenko},\ and\ \citenamefont {Takhistov}}]{Cotner:2018vug}%
  \BibitemOpen
  \bibfield  {author} {\bibinfo {author} {\bibfnamefont {E.}~\bibnamefont
  {Cotner}}, \bibinfo {author} {\bibfnamefont {A.}~\bibnamefont {Kusenko}}, \
  and\ \bibinfo {author} {\bibfnamefont {V.}~\bibnamefont {Takhistov}},\ }\href
  {\doibase 10.1103/PhysRevD.98.083513} {\bibfield  {journal} {\bibinfo
  {journal} {Phys. Rev.}\ }\textbf {\bibinfo {volume} {D98}},\ \bibinfo {pages}
  {083513} (\bibinfo {year} {2018})},\ \Eprint
  {http://arxiv.org/abs/1801.03321} {arXiv:1801.03321 [astro-ph.CO]}
  \BibitemShut {NoStop}%
%%CITATION = ARXIV:1801.03321;%%
\bibitem [{\citenamefont {Gregory}\ \emph {et~al.}(2017)\citenamefont
  {Gregory}, \citenamefont {Kastor},\ and\ \citenamefont
  {Traschen}}]{Gregory:2017sor}%
  \BibitemOpen
  \bibfield  {author} {\bibinfo {author} {\bibfnamefont {R.}~\bibnamefont
  {Gregory}}, \bibinfo {author} {\bibfnamefont {D.}~\bibnamefont {Kastor}}, \
  and\ \bibinfo {author} {\bibfnamefont {J.}~\bibnamefont {Traschen}},\ }\href
  {\doibase 10.1007/JHEP10(2017)118} {\bibfield  {journal} {\bibinfo  {journal}
  {JHEP}\ }\textbf {\bibinfo {volume} {10}},\ \bibinfo {pages} {118} (\bibinfo
  {year} {2017})},\ \Eprint {http://arxiv.org/abs/1707.06586} {arXiv:1707.06586
  [hep-th]} \BibitemShut {NoStop}%
%%CITATION = ARXIV:1707.06586;%%
\bibitem [{\citenamefont {Carr}\ \emph
  {et~al.}(2017{\natexlab{a}})\citenamefont {Carr}, \citenamefont {Tenkanen},\
  and\ \citenamefont {Vaskonen}}]{Carr:2017edp}%
  \BibitemOpen
  \bibfield  {author} {\bibinfo {author} {\bibfnamefont {B.}~\bibnamefont
  {Carr}}, \bibinfo {author} {\bibfnamefont {T.}~\bibnamefont {Tenkanen}}, \
  and\ \bibinfo {author} {\bibfnamefont {V.}~\bibnamefont {Vaskonen}},\ }\href
  {\doibase 10.1103/PhysRevD.96.063507} {\bibfield  {journal} {\bibinfo
  {journal} {Phys. Rev.}\ }\textbf {\bibinfo {volume} {D96}},\ \bibinfo {pages}
  {063507} (\bibinfo {year} {2017}{\natexlab{a}})},\ \Eprint
  {http://arxiv.org/abs/1706.03746} {arXiv:1706.03746 [astro-ph.CO]}
  \BibitemShut {NoStop}%
%%CITATION = ARXIV:1706.03746;%%
\bibitem [{\citenamefont {Dolgov}(2018)}]{Dolgov:2017aec}%
  \BibitemOpen
  \bibfield  {author} {\bibinfo {author} {\bibfnamefont {A.~D.}\ \bibnamefont
  {Dolgov}},\ }\href {\doibase 10.3367/UFNe.2017.06.038153,
  10.3367/UFNr.2017.06.038153} {\bibfield  {journal} {\bibinfo  {journal} {Usp.
  Fiz. Nauk}\ }\textbf {\bibinfo {volume} {188}},\ \bibinfo {pages} {121}
  (\bibinfo {year} {2018})},\ \bibinfo {note} {[Phys. Usp.61,no.2,115(2018)]},\
  \Eprint {http://arxiv.org/abs/1705.06859} {arXiv:1705.06859 [astro-ph.CO]}
  \BibitemShut {NoStop}%
%%CITATION = ARXIV:1705.06859;%%
\bibitem [{\citenamefont {Domcke}\ \emph {et~al.}(2017)\citenamefont {Domcke},
  \citenamefont {Muia}, \citenamefont {Pieroni},\ and\ \citenamefont
  {Witkowski}}]{Domcke:2017fix}%
  \BibitemOpen
  \bibfield  {author} {\bibinfo {author} {\bibfnamefont {V.}~\bibnamefont
  {Domcke}}, \bibinfo {author} {\bibfnamefont {F.}~\bibnamefont {Muia}},
  \bibinfo {author} {\bibfnamefont {M.}~\bibnamefont {Pieroni}}, \ and\
  \bibinfo {author} {\bibfnamefont {L.~T.}\ \bibnamefont {Witkowski}},\ }\href
  {\doibase 10.1088/1475-7516/2017/07/048} {\bibfield  {journal} {\bibinfo
  {journal} {JCAP}\ }\textbf {\bibinfo {volume} {1707}},\ \bibinfo {pages}
  {048} (\bibinfo {year} {2017})},\ \Eprint {http://arxiv.org/abs/1704.03464}
  {arXiv:1704.03464 [astro-ph.CO]} \BibitemShut {NoStop}%
%%CITATION = ARXIV:1704.03464;%%
\bibitem [{\citenamefont {Soni}\ \emph {et~al.}(2017)\citenamefont {Soni},
  \citenamefont {Xiao},\ and\ \citenamefont {Zhang}}]{Soni:2017nlm}%
  \BibitemOpen
  \bibfield  {author} {\bibinfo {author} {\bibfnamefont {A.}~\bibnamefont
  {Soni}}, \bibinfo {author} {\bibfnamefont {H.}~\bibnamefont {Xiao}}, \ and\
  \bibinfo {author} {\bibfnamefont {Y.}~\bibnamefont {Zhang}},\ }\href
  {\doibase 10.1103/PhysRevD.96.083514} {\bibfield  {journal} {\bibinfo
  {journal} {Phys. Rev.}\ }\textbf {\bibinfo {volume} {D96}},\ \bibinfo {pages}
  {083514} (\bibinfo {year} {2017})},\ \Eprint
  {http://arxiv.org/abs/1704.02347} {arXiv:1704.02347 [hep-ph]} \BibitemShut
  {NoStop}%
%%CITATION = ARXIV:1704.02347;%%
\bibitem [{\citenamefont {Quintin}\ and\ \citenamefont
  {Brandenberger}(2016)}]{Quintin:2016qro}%
  \BibitemOpen
  \bibfield  {author} {\bibinfo {author} {\bibfnamefont {J.}~\bibnamefont
  {Quintin}}\ and\ \bibinfo {author} {\bibfnamefont {R.~H.}\ \bibnamefont
  {Brandenberger}},\ }\href {\doibase 10.1088/1475-7516/2016/11/029} {\bibfield
   {journal} {\bibinfo  {journal} {JCAP}\ }\textbf {\bibinfo {volume} {1611}},\
  \bibinfo {pages} {029} (\bibinfo {year} {2016})},\ \Eprint
  {http://arxiv.org/abs/1609.02556} {arXiv:1609.02556 [astro-ph.CO]}
  \BibitemShut {NoStop}%
%%CITATION = ARXIV:1609.02556;%%
\bibitem [{\citenamefont {Belotsky}\ \emph {et~al.}(2018)\citenamefont
  {Belotsky}, \citenamefont {Dokuchaev}, \citenamefont {Eroshenko},
  \citenamefont {Esipova}, \citenamefont {Khlopov}, \citenamefont {Khromykh},
  \citenamefont {Kirillov}, \citenamefont {Nikulin}, \citenamefont {Rubin},\
  and\ \citenamefont {Svadkovsky}}]{Belotsky:2018wph}%
  \BibitemOpen
  \bibfield  {author} {\bibinfo {author} {\bibfnamefont {K.~M.}\ \bibnamefont
  {Belotsky}}, \bibinfo {author} {\bibfnamefont {V.~I.}\ \bibnamefont
  {Dokuchaev}}, \bibinfo {author} {\bibfnamefont {Y.~N.}\ \bibnamefont
  {Eroshenko}}, \bibinfo {author} {\bibfnamefont {E.~A.}\ \bibnamefont
  {Esipova}}, \bibinfo {author} {\bibfnamefont {M.~{\relax Yu}.}\ \bibnamefont
  {Khlopov}}, \bibinfo {author} {\bibfnamefont {L.~A.}\ \bibnamefont
  {Khromykh}}, \bibinfo {author} {\bibfnamefont {A.~A.}\ \bibnamefont
  {Kirillov}}, \bibinfo {author} {\bibfnamefont {V.~V.}\ \bibnamefont
  {Nikulin}}, \bibinfo {author} {\bibfnamefont {S.~G.}\ \bibnamefont {Rubin}},
  \ and\ \bibinfo {author} {\bibfnamefont {I.~V.}\ \bibnamefont {Svadkovsky}},\
  }\href@noop {} {\  (\bibinfo {year} {2018})},\ \Eprint
  {http://arxiv.org/abs/1807.06590} {arXiv:1807.06590 [astro-ph.CO]}
  \BibitemShut {NoStop}%
%%CITATION = ARXIV:1807.06590;%%
\bibitem [{\citenamefont {Clark}\ \emph {et~al.}(2017)\citenamefont {Clark},
  \citenamefont {Dutta}, \citenamefont {Gao}, \citenamefont {Strigari},\ and\
  \citenamefont {Watson}}]{Clark:2016nst}%
  \BibitemOpen
  \bibfield  {author} {\bibinfo {author} {\bibfnamefont {S.}~\bibnamefont
  {Clark}}, \bibinfo {author} {\bibfnamefont {B.}~\bibnamefont {Dutta}},
  \bibinfo {author} {\bibfnamefont {Y.}~\bibnamefont {Gao}}, \bibinfo {author}
  {\bibfnamefont {L.~E.}\ \bibnamefont {Strigari}}, \ and\ \bibinfo {author}
  {\bibfnamefont {S.}~\bibnamefont {Watson}},\ }\href {\doibase
  10.1103/PhysRevD.95.083006} {\bibfield  {journal} {\bibinfo  {journal} {Phys.
  Rev.}\ }\textbf {\bibinfo {volume} {D95}},\ \bibinfo {pages} {083006}
  (\bibinfo {year} {2017})},\ \Eprint {http://arxiv.org/abs/1612.07738}
  {arXiv:1612.07738 [astro-ph.CO]} \BibitemShut {NoStop}%
%%CITATION = ARXIV:1612.07738;%%
\bibitem [{\citenamefont {Dalianis}(2018)}]{Dalianis:2018ymb}%
  \BibitemOpen
  \bibfield  {author} {\bibinfo {author} {\bibfnamefont {I.}~\bibnamefont
  {Dalianis}},\ }\href@noop {} {\  (\bibinfo {year} {2018})},\ \Eprint
  {http://arxiv.org/abs/1812.09807} {arXiv:1812.09807 [astro-ph.CO]}
  \BibitemShut {NoStop}%
%%CITATION = ARXIV:1812.09807;%%
\bibitem [{\citenamefont {Clark}\ \emph {et~al.}(2018)\citenamefont {Clark},
  \citenamefont {Dutta}, \citenamefont {Gao}, \citenamefont {Ma},\ and\
  \citenamefont {Strigari}}]{Clark:2018ghm}%
  \BibitemOpen
  \bibfield  {author} {\bibinfo {author} {\bibfnamefont {S.}~\bibnamefont
  {Clark}}, \bibinfo {author} {\bibfnamefont {B.}~\bibnamefont {Dutta}},
  \bibinfo {author} {\bibfnamefont {Y.}~\bibnamefont {Gao}}, \bibinfo {author}
  {\bibfnamefont {Y.-Z.}\ \bibnamefont {Ma}}, \ and\ \bibinfo {author}
  {\bibfnamefont {L.~E.}\ \bibnamefont {Strigari}},\ }\href {\doibase
  10.1103/PhysRevD.98.043006} {\bibfield  {journal} {\bibinfo  {journal} {Phys.
  Rev.}\ }\textbf {\bibinfo {volume} {D98}},\ \bibinfo {pages} {043006}
  (\bibinfo {year} {2018})},\ \Eprint {http://arxiv.org/abs/1803.09390}
  {arXiv:1803.09390 [astro-ph.HE]} \BibitemShut {NoStop}%
%%CITATION = ARXIV:1803.09390;%%
\bibitem [{\citenamefont {Takhistov}(2019)}]{Takhistov:2017nmt}%
  \BibitemOpen
  \bibfield  {author} {\bibinfo {author} {\bibfnamefont {V.}~\bibnamefont
  {Takhistov}},\ }\href {\doibase 10.1016/j.physletb.2018.12.043} {\bibfield
  {journal} {\bibinfo  {journal} {Phys. Lett.}\ }\textbf {\bibinfo {volume}
  {B789}},\ \bibinfo {pages} {538} (\bibinfo {year} {2019})},\ \Eprint
  {http://arxiv.org/abs/1710.09458} {arXiv:1710.09458 [astro-ph.HE]}
  \BibitemShut {NoStop}%
%%CITATION = ARXIV:1710.09458;%%
\bibitem [{\citenamefont {Guo}\ \emph {et~al.}(2019)\citenamefont {Guo},
  \citenamefont {Shu},\ and\ \citenamefont {Zhao}}]{Guo:2017njn}%
  \BibitemOpen
  \bibfield  {author} {\bibinfo {author} {\bibfnamefont {H.-K.}\ \bibnamefont
  {Guo}}, \bibinfo {author} {\bibfnamefont {J.}~\bibnamefont {Shu}}, \ and\
  \bibinfo {author} {\bibfnamefont {Y.}~\bibnamefont {Zhao}},\ }\href {\doibase
  10.1103/PhysRevD.99.023001} {\bibfield  {journal} {\bibinfo  {journal} {Phys.
  Rev.}\ }\textbf {\bibinfo {volume} {D99}},\ \bibinfo {pages} {023001}
  (\bibinfo {year} {2019})},\ \Eprint {http://arxiv.org/abs/1709.03500}
  {arXiv:1709.03500 [astro-ph.CO]} \BibitemShut {NoStop}%
%%CITATION = ARXIV:1709.03500;%%
\bibitem [{\citenamefont {Takhistov}(2018)}]{Takhistov:2017bpt}%
  \BibitemOpen
  \bibfield  {author} {\bibinfo {author} {\bibfnamefont {V.}~\bibnamefont
  {Takhistov}},\ }\href {\doibase 10.1016/j.physletb.2018.05.026} {\bibfield
  {journal} {\bibinfo  {journal} {Phys. Lett.}\ }\textbf {\bibinfo {volume}
  {B782}},\ \bibinfo {pages} {77} (\bibinfo {year} {2018})},\ \Eprint
  {http://arxiv.org/abs/1707.05849} {arXiv:1707.05849 [astro-ph.CO]}
  \BibitemShut {NoStop}%
%%CITATION = ARXIV:1707.05849;%%
\bibitem [{\citenamefont {Cole}\ and\ \citenamefont
  {Byrnes}(2018)}]{Cole:2017gle}%
  \BibitemOpen
  \bibfield  {author} {\bibinfo {author} {\bibfnamefont {P.~S.}\ \bibnamefont
  {Cole}}\ and\ \bibinfo {author} {\bibfnamefont {C.~T.}\ \bibnamefont
  {Byrnes}},\ }\href {\doibase 10.1088/1475-7516/2018/02/019} {\bibfield
  {journal} {\bibinfo  {journal} {JCAP}\ }\textbf {\bibinfo {volume} {1802}},\
  \bibinfo {pages} {019} (\bibinfo {year} {2018})},\ \Eprint
  {http://arxiv.org/abs/1706.10288} {arXiv:1706.10288 [astro-ph.CO]}
  \BibitemShut {NoStop}%
%%CITATION = ARXIV:1706.10288;%%
\bibitem [{\citenamefont {Emami}\ and\ \citenamefont
  {Smoot}(2018)}]{Emami:2017fiy}%
  \BibitemOpen
  \bibfield  {author} {\bibinfo {author} {\bibfnamefont {R.}~\bibnamefont
  {Emami}}\ and\ \bibinfo {author} {\bibfnamefont {G.}~\bibnamefont {Smoot}},\
  }\href {\doibase 10.1088/1475-7516/2018/01/007} {\bibfield  {journal}
  {\bibinfo  {journal} {JCAP}\ }\textbf {\bibinfo {volume} {1801}},\ \bibinfo
  {pages} {007} (\bibinfo {year} {2018})},\ \Eprint
  {http://arxiv.org/abs/1705.09924} {arXiv:1705.09924 [astro-ph.CO]}
  \BibitemShut {NoStop}%
%%CITATION = ARXIV:1705.09924;%%
\bibitem [{\citenamefont {Inoue}\ and\ \citenamefont
  {Kusenko}(2017)}]{Inoue:2017csr}%
  \BibitemOpen
  \bibfield  {author} {\bibinfo {author} {\bibfnamefont {Y.}~\bibnamefont
  {Inoue}}\ and\ \bibinfo {author} {\bibfnamefont {A.}~\bibnamefont
  {Kusenko}},\ }\href {\doibase 10.1088/1475-7516/2017/10/034} {\bibfield
  {journal} {\bibinfo  {journal} {JCAP}\ }\textbf {\bibinfo {volume} {1710}},\
  \bibinfo {pages} {034} (\bibinfo {year} {2017})},\ \Eprint
  {http://arxiv.org/abs/1705.00791} {arXiv:1705.00791 [astro-ph.CO]}
  \BibitemShut {NoStop}%
%%CITATION = ARXIV:1705.00791;%%
\bibitem [{\citenamefont {Gong}\ and\ \citenamefont
  {Kitajima}(2017)}]{Gong:2017sie}%
  \BibitemOpen
  \bibfield  {author} {\bibinfo {author} {\bibfnamefont {J.-O.}\ \bibnamefont
  {Gong}}\ and\ \bibinfo {author} {\bibfnamefont {N.}~\bibnamefont
  {Kitajima}},\ }\href {\doibase 10.1088/1475-7516/2017/08/017} {\bibfield
  {journal} {\bibinfo  {journal} {JCAP}\ }\textbf {\bibinfo {volume} {1708}},\
  \bibinfo {pages} {017} (\bibinfo {year} {2017})},\ \Eprint
  {http://arxiv.org/abs/1704.04132} {arXiv:1704.04132 [astro-ph.CO]}
  \BibitemShut {NoStop}%
%%CITATION = ARXIV:1704.04132;%%
\bibitem [{\citenamefont {Lehmann}\ \emph {et~al.}(2018)\citenamefont
  {Lehmann}, \citenamefont {Profumo},\ and\ \citenamefont
  {Yant}}]{Lehmann2018}%
  \BibitemOpen
  \bibfield  {author} {\bibinfo {author} {\bibfnamefont {B.~V.}\ \bibnamefont
  {Lehmann}}, \bibinfo {author} {\bibfnamefont {S.}~\bibnamefont {Profumo}}, \
  and\ \bibinfo {author} {\bibfnamefont {J.}~\bibnamefont {Yant}},\ }\href
  {\doibase 10.1088/1475-7516/2018/04/007} {\bibfield  {journal} {\bibinfo
  {journal} {JCAP}\ }\textbf {\bibinfo {volume} {1804}},\ \bibinfo {pages}
  {007} (\bibinfo {year} {2018})},\ \Eprint {http://arxiv.org/abs/1801.00808}
  {arXiv:1801.00808 [astro-ph.CO]} \BibitemShut {NoStop}%
%%CITATION = ARXIV:1801.00808;%%
\bibitem [{\citenamefont {Dror}\ \emph {et~al.}(2019)\citenamefont {Dror},
  \citenamefont {Ramani}, \citenamefont {Trickle},\ and\ \citenamefont
  {Zurek}}]{Dror:2019twh}%
  \BibitemOpen
  \bibfield  {author} {\bibinfo {author} {\bibfnamefont {J.~A.}\ \bibnamefont
  {Dror}}, \bibinfo {author} {\bibfnamefont {H.}~\bibnamefont {Ramani}},
  \bibinfo {author} {\bibfnamefont {T.}~\bibnamefont {Trickle}}, \ and\
  \bibinfo {author} {\bibfnamefont {K.~M.}\ \bibnamefont {Zurek}},\ }\href@noop
  {} {\  (\bibinfo {year} {2019})},\ \Eprint {http://arxiv.org/abs/1901.04490}
  {arXiv:1901.04490 [astro-ph.CO]} \BibitemShut {NoStop}%
%%CITATION = ARXIV:1901.04490;%%
\bibitem [{\citenamefont {{Novikov}}(1975)}]{Nov1975}%
  \BibitemOpen
  \bibfield  {author} {\bibinfo {author} {\bibfnamefont {I.~D.}\ \bibnamefont
  {{Novikov}}},\ }\href@noop {} {\bibfield  {journal} {\bibinfo  {journal}
  {Astronomicheskii Zhurnal}\ }\textbf {\bibinfo {volume} {52}},\ \bibinfo
  {pages} {1038} (\bibinfo {year} {1975})}\BibitemShut {NoStop}%
\bibitem [{\citenamefont {Zeldovich}(1970)}]{Zeldovich:1969sb}%
  \BibitemOpen
  \bibfield  {author} {\bibinfo {author} {\bibfnamefont {{\relax Ya}.~B.}\
  \bibnamefont {Zeldovich}},\ }\href@noop {} {\bibfield  {journal} {\bibinfo
  {journal} {Astron. Astrophys.}\ }\textbf {\bibinfo {volume} {5}},\ \bibinfo
  {pages} {84} (\bibinfo {year} {1970})}\BibitemShut {NoStop}%
%%CITATION = AAEJA,5,84;%%
\bibitem [{\citenamefont {{Zeldovich}}\ and\ \citenamefont
  {{Novikov}}(1983)}]{Zel:1983}%
  \BibitemOpen
  \bibfield  {author} {\bibinfo {author} {\bibfnamefont {I.~B.}\ \bibnamefont
  {{Zeldovich}}}\ and\ \bibinfo {author} {\bibfnamefont {I.~D.}\ \bibnamefont
  {{Novikov}}},\ }\href@noop {} {\emph {\bibinfo {title} {{Relativistic
  astrophysics. Volume 2.}}}}\ (\bibinfo  {publisher} {Chicago, IL, University
  of Chicago Press, 1983, 751 p.~Translation.},\ \bibinfo {year}
  {1983})\BibitemShut {NoStop}%
\bibitem [{\citenamefont {{Doroshkevich}}(1970)}]{Dorosh1970}%
  \BibitemOpen
  \bibfield  {author} {\bibinfo {author} {\bibfnamefont {A.~G.}\ \bibnamefont
  {{Doroshkevich}}},\ }\href@noop {} {\bibfield  {journal} {\bibinfo  {journal}
  {Astrofizika}\ }\textbf {\bibinfo {volume} {6}},\ \bibinfo {pages} {581}
  (\bibinfo {year} {1970})}\BibitemShut {NoStop}%
\bibitem [{\citenamefont {{Doroshkevich}}\ and\ \citenamefont
  {{Shandarin}}(1978)}]{Dorosh1978}%
  \BibitemOpen
  \bibfield  {author} {\bibinfo {author} {\bibfnamefont {A.~G.}\ \bibnamefont
  {{Doroshkevich}}}\ and\ \bibinfo {author} {\bibfnamefont {S.~F.}\
  \bibnamefont {{Shandarin}}},\ }\href@noop {} {\bibfield  {journal} {\bibinfo
  {journal} {Astronomicheskii Zhurnal}\ }\textbf {\bibinfo {volume} {22}},\
  \bibinfo {pages} {653} (\bibinfo {year} {1978})}\BibitemShut {NoStop}%
\bibitem [{\citenamefont {{Polnarev}}\ and\ \citenamefont
  {{Khlopov}}(1981)}]{KP81}%
  \BibitemOpen
  \bibfield  {author} {\bibinfo {author} {\bibfnamefont {A.~G.}\ \bibnamefont
  {{Polnarev}}}\ and\ \bibinfo {author} {\bibfnamefont {M.~Y.}\ \bibnamefont
  {{Khlopov}}},\ }\href@noop {} {\bibfield  {journal} {\bibinfo  {journal}
  {Astronomicheskii Zhurnal}\ }\textbf {\bibinfo {volume} {25}},\ \bibinfo
  {pages} {406} (\bibinfo {year} {1981})}\BibitemShut {NoStop}%
\bibitem [{\citenamefont {Kokubu}\ \emph {et~al.}(2018)\citenamefont {Kokubu},
  \citenamefont {Kyutoku}, \citenamefont {Kohri},\ and\ \citenamefont
  {Harada}}]{Kokubu2018}%
  \BibitemOpen
  \bibfield  {author} {\bibinfo {author} {\bibfnamefont {T.}~\bibnamefont
  {Kokubu}}, \bibinfo {author} {\bibfnamefont {K.}~\bibnamefont {Kyutoku}},
  \bibinfo {author} {\bibfnamefont {K.}~\bibnamefont {Kohri}}, \ and\ \bibinfo
  {author} {\bibfnamefont {T.}~\bibnamefont {Harada}},\ }\href@noop {} {\
  (\bibinfo {year} {2018})},\ \Eprint {http://arxiv.org/abs/1810.03490}
  {arXiv:1810.03490 [astro-ph.CO]} \BibitemShut {NoStop}%
%%CITATION = ARXIV:1810.03490;%%
\bibitem [{\citenamefont {Harada}\ \emph {et~al.}(2016)\citenamefont {Harada},
  \citenamefont {Yoo}, \citenamefont {Kohri}, \citenamefont {Nakao},\ and\
  \citenamefont {Jhingan}}]{Harada2016}%
  \BibitemOpen
  \bibfield  {author} {\bibinfo {author} {\bibfnamefont {T.}~\bibnamefont
  {Harada}}, \bibinfo {author} {\bibfnamefont {C.-M.}\ \bibnamefont {Yoo}},
  \bibinfo {author} {\bibfnamefont {K.}~\bibnamefont {Kohri}}, \bibinfo
  {author} {\bibfnamefont {K.-i.}\ \bibnamefont {Nakao}}, \ and\ \bibinfo
  {author} {\bibfnamefont {S.}~\bibnamefont {Jhingan}},\ }\href {\doibase
  10.3847/1538-4357/833/1/61} {\bibfield  {journal} {\bibinfo  {journal}
  {Astrophys. J.}\ }\textbf {\bibinfo {volume} {833}},\ \bibinfo {pages} {61}
  (\bibinfo {year} {2016})},\ \Eprint {http://arxiv.org/abs/1609.01588}
  {arXiv:1609.01588 [astro-ph.CO]} \BibitemShut {NoStop}%
%%CITATION = ARXIV:1609.01588;%%
\bibitem [{\citenamefont {Malec}\ and\ \citenamefont {Xie}(2015)}]{Malec2015}%
  \BibitemOpen
  \bibfield  {author} {\bibinfo {author} {\bibfnamefont {E.}~\bibnamefont
  {Malec}}\ and\ \bibinfo {author} {\bibfnamefont {N.}~\bibnamefont {Xie}},\
  }\href {\doibase 10.1103/PhysRevD.91.081501} {\bibfield  {journal} {\bibinfo
  {journal} {Phys. Rev.}\ }\textbf {\bibinfo {volume} {D91}},\ \bibinfo {pages}
  {081501} (\bibinfo {year} {2015})},\ \Eprint
  {http://arxiv.org/abs/1503.01354} {arXiv:1503.01354 [gr-qc]} \BibitemShut
  {NoStop}%
%%CITATION = ARXIV:1503.01354;%%
\bibitem [{\citenamefont {{Coles}}\ \emph {et~al.}(1993)\citenamefont
  {{Coles}}, \citenamefont {{Melott}},\ and\ \citenamefont
  {{Shandarin}}}]{coles}%
  \BibitemOpen
  \bibfield  {author} {\bibinfo {author} {\bibfnamefont {P.}~\bibnamefont
  {{Coles}}}, \bibinfo {author} {\bibfnamefont {A.~L.}\ \bibnamefont
  {{Melott}}}, \ and\ \bibinfo {author} {\bibfnamefont {S.~F.}\ \bibnamefont
  {{Shandarin}}},\ }\href {\doibase 10.1093/mnras/260.4.765} {\bibfield
  {journal} {\bibinfo  {journal} {Monthly Notices of the Royal Astronomical
  Society}\ }\textbf {\bibinfo {volume} {260}},\ \bibinfo {pages} {765}
  (\bibinfo {year} {1993})}\BibitemShut {NoStop}%
\bibitem [{\citenamefont {Brandenberger}(2004)}]{Brandenberger:2003vk}%
  \BibitemOpen
  \bibfield  {author} {\bibinfo {author} {\bibfnamefont {R.~H.}\ \bibnamefont
  {Brandenberger}},\ }\href {\doibase 10.1007/978-3-540-40918-2_5} {\bibfield
  {journal} {\bibinfo  {journal} {Lect. Notes Phys.}\ }\textbf {\bibinfo
  {volume} {646}},\ \bibinfo {pages} {127} (\bibinfo {year} {2004})},\ \Eprint
  {http://arxiv.org/abs/hep-th/0306071} {arXiv:hep-th/0306071 [hep-th]}
  \BibitemShut {NoStop}%
%%CITATION = HEP-TH/0306071;%%
\bibitem [{\citenamefont {Goldstein}\ \emph {et~al.}(2002)\citenamefont
  {Goldstein}, \citenamefont {Poole},\ and\ \citenamefont
  {Safko}}]{goldstein2002classical}%
  \BibitemOpen
  \bibfield  {author} {\bibinfo {author} {\bibfnamefont {H.}~\bibnamefont
  {Goldstein}}, \bibinfo {author} {\bibfnamefont {C.}~\bibnamefont {Poole}}, \
  and\ \bibinfo {author} {\bibfnamefont {J.}~\bibnamefont {Safko}},\ }\href
  {https://books.google.com/books?id=tJCuQgAACAAJ} {\emph {\bibinfo {title}
  {Classical Mechanics}}}\ (\bibinfo  {publisher} {Addison Wesley},\ \bibinfo
  {year} {2002})\BibitemShut {NoStop}%
\bibitem [{\citenamefont {Misner}\ \emph {et~al.}(1973)\citenamefont {Misner},
  \citenamefont {Thorne},\ and\ \citenamefont {Wheeler}}]{Misner1974}%
  \BibitemOpen
  \bibfield  {author} {\bibinfo {author} {\bibfnamefont {C.~W.}\ \bibnamefont
  {Misner}}, \bibinfo {author} {\bibfnamefont {K.~S.}\ \bibnamefont {Thorne}},
  \ and\ \bibinfo {author} {\bibfnamefont {J.~A.}\ \bibnamefont {Wheeler}},\
  }\href@noop {} {\emph {\bibinfo {title} {{Gravitation}}}}\ (\bibinfo
  {publisher} {W. H. Freeman},\ \bibinfo {address} {San Francisco},\ \bibinfo
  {year} {1973})\BibitemShut {NoStop}%
%%CITATION = INSPIRE-95654;%%
\bibitem [{\citenamefont {Lemaitre}(1997)}]{LemTB}%
  \BibitemOpen
  \bibfield  {author} {\bibinfo {author} {\bibfnamefont {G.}~\bibnamefont
  {Lemaitre}},\ }\href {\doibase 10.1023/A:1018855621348} {\bibfield  {journal}
  {\bibinfo  {journal} {Gen. Rel. Grav.}\ }\textbf {\bibinfo {volume} {29}},\
  \bibinfo {pages} {641} (\bibinfo {year} {1997})},\ \bibinfo {note} {[Annales
  Soc. Sci. Bruxelles A53,51(1933)]}\BibitemShut {NoStop}%
%%CITATION = GRGVA,29,641;%%
\bibitem [{\citenamefont {Tolman}(1934)}]{LTolB}%
  \BibitemOpen
  \bibfield  {author} {\bibinfo {author} {\bibfnamefont {R.~C.}\ \bibnamefont
  {Tolman}},\ }\href {\doibase 10.1073/pnas.20.3.169} {\bibfield  {journal}
  {\bibinfo  {journal} {Proc. Nat. Acad. Sci.}\ }\textbf {\bibinfo {volume}
  {20}},\ \bibinfo {pages} {169} (\bibinfo {year} {1934})},\ \bibinfo {note}
  {[Gen. Rel. Grav.29,935(1997)]}\BibitemShut {NoStop}%
%%CITATION = PNASA,20,169;%%
\bibitem [{\citenamefont {Bondi}(1947)}]{LTBond}%
  \BibitemOpen
  \bibfield  {author} {\bibinfo {author} {\bibfnamefont {H.}~\bibnamefont
  {Bondi}},\ }\href {\doibase 10.1093/mnras/107.5-6.410} {\bibfield  {journal}
  {\bibinfo  {journal} {Mon. Not. Roy. Astron. Soc.}\ }\textbf {\bibinfo
  {volume} {107}},\ \bibinfo {pages} {410} (\bibinfo {year}
  {1947})}\BibitemShut {NoStop}%
%%CITATION = MNRAA,107,410;%%
\bibitem [{\citenamefont {Carr}\ \emph {et~al.}(2010)\citenamefont {Carr},
  \citenamefont {Kohri}, \citenamefont {Sendouda},\ and\ \citenamefont
  {Yokoyama}}]{Carr:2009jm}%
  \BibitemOpen
  \bibfield  {author} {\bibinfo {author} {\bibfnamefont {B.~J.}\ \bibnamefont
  {Carr}}, \bibinfo {author} {\bibfnamefont {K.}~\bibnamefont {Kohri}},
  \bibinfo {author} {\bibfnamefont {Y.}~\bibnamefont {Sendouda}}, \ and\
  \bibinfo {author} {\bibfnamefont {J.}~\bibnamefont {Yokoyama}},\ }\href
  {\doibase 10.1103/PhysRevD.81.104019} {\bibfield  {journal} {\bibinfo
  {journal} {Phys. Rev.}\ }\textbf {\bibinfo {volume} {D81}},\ \bibinfo {pages}
  {104019} (\bibinfo {year} {2010})},\ \Eprint {http://arxiv.org/abs/0912.5297}
  {arXiv:0912.5297 [astro-ph.CO]} \BibitemShut {NoStop}%
%%CITATION = ARXIV:0912.5297;%%
\bibitem [{\citenamefont {Barnacka}\ \emph {et~al.}(2012)\citenamefont
  {Barnacka}, \citenamefont {Glicenstein},\ and\ \citenamefont
  {Moderski}}]{Barnacka:2012bm}%
  \BibitemOpen
  \bibfield  {author} {\bibinfo {author} {\bibfnamefont {A.}~\bibnamefont
  {Barnacka}}, \bibinfo {author} {\bibfnamefont {J.~F.}\ \bibnamefont
  {Glicenstein}}, \ and\ \bibinfo {author} {\bibfnamefont {R.}~\bibnamefont
  {Moderski}},\ }\href {\doibase 10.1103/PhysRevD.86.043001} {\bibfield
  {journal} {\bibinfo  {journal} {Phys. Rev.}\ }\textbf {\bibinfo {volume}
  {D86}},\ \bibinfo {pages} {043001} (\bibinfo {year} {2012})},\ \Eprint
  {http://arxiv.org/abs/1204.2056} {arXiv:1204.2056 [astro-ph.CO]} \BibitemShut
  {NoStop}%
%%CITATION = ARXIV:1204.2056;%%
\bibitem [{\citenamefont {Graham}\ \emph {et~al.}(2015)\citenamefont {Graham},
  \citenamefont {Rajendran},\ and\ \citenamefont {Varela}}]{Graham:2015apa}%
  \BibitemOpen
  \bibfield  {author} {\bibinfo {author} {\bibfnamefont {P.~W.}\ \bibnamefont
  {Graham}}, \bibinfo {author} {\bibfnamefont {S.}~\bibnamefont {Rajendran}}, \
  and\ \bibinfo {author} {\bibfnamefont {J.}~\bibnamefont {Varela}},\ }\href
  {\doibase 10.1103/PhysRevD.92.063007} {\bibfield  {journal} {\bibinfo
  {journal} {Phys. Rev.}\ }\textbf {\bibinfo {volume} {D92}},\ \bibinfo {pages}
  {063007} (\bibinfo {year} {2015})},\ \Eprint
  {http://arxiv.org/abs/1505.04444} {arXiv:1505.04444 [hep-ph]} \BibitemShut
  {NoStop}%
%%CITATION = ARXIV:1505.04444;%%
\bibitem [{\citenamefont {Niikura}\ \emph {et~al.}(2017)\citenamefont {Niikura}
  \emph {et~al.}}]{Niikura:2017zjd}%
  \BibitemOpen
  \bibfield  {author} {\bibinfo {author} {\bibfnamefont {H.}~\bibnamefont
  {Niikura}} \emph {et~al.},\ }\href@noop {} {\  (\bibinfo {year} {2017})},\
  \Eprint {http://arxiv.org/abs/1701.02151} {arXiv:1701.02151 [astro-ph.CO]}
  \BibitemShut {NoStop}%
%%CITATION = ARXIV:1701.02151;%%
\bibitem [{\citenamefont {Griest}\ \emph {et~al.}(2014)\citenamefont {Griest},
  \citenamefont {Cieplak},\ and\ \citenamefont {Lehner}}]{Griest:2013aaa}%
  \BibitemOpen
  \bibfield  {author} {\bibinfo {author} {\bibfnamefont {K.}~\bibnamefont
  {Griest}}, \bibinfo {author} {\bibfnamefont {A.~M.}\ \bibnamefont {Cieplak}},
  \ and\ \bibinfo {author} {\bibfnamefont {M.~J.}\ \bibnamefont {Lehner}},\
  }\href {\doibase 10.1088/0004-637X/786/2/158} {\bibfield  {journal} {\bibinfo
   {journal} {Astrophys. J.}\ }\textbf {\bibinfo {volume} {786}},\ \bibinfo
  {pages} {158} (\bibinfo {year} {2014})},\ \Eprint
  {http://arxiv.org/abs/1307.5798} {arXiv:1307.5798 [astro-ph.CO]} \BibitemShut
  {NoStop}%
%%CITATION = ARXIV:1307.5798;%%
\bibitem [{\citenamefont {Tisserand}\ \emph {et~al.}(2007)\citenamefont
  {Tisserand} \emph {et~al.}}]{Tisserand:2006zx}%
  \BibitemOpen
  \bibfield  {author} {\bibinfo {author} {\bibfnamefont {P.}~\bibnamefont
  {Tisserand}} \emph {et~al.} (\bibinfo {collaboration} {EROS-2}),\ }\href
  {\doibase 10.1051/0004-6361:20066017} {\bibfield  {journal} {\bibinfo
  {journal} {Astron. Astrophys.}\ }\textbf {\bibinfo {volume} {469}},\ \bibinfo
  {pages} {387} (\bibinfo {year} {2007})},\ \Eprint
  {http://arxiv.org/abs/astro-ph/0607207} {arXiv:astro-ph/0607207 [astro-ph]}
  \BibitemShut {NoStop}%
%%CITATION = ASTRO-PH/0607207;%%
\bibitem [{\citenamefont {Zumalacarregui}\ and\ \citenamefont
  {Seljak}(2018)}]{Zumalacarregui:2017qqd}%
  \BibitemOpen
  \bibfield  {author} {\bibinfo {author} {\bibfnamefont {M.}~\bibnamefont
  {Zumalacarregui}}\ and\ \bibinfo {author} {\bibfnamefont {U.}~\bibnamefont
  {Seljak}},\ }\href {\doibase 10.1103/PhysRevLett.121.141101} {\bibfield
  {journal} {\bibinfo  {journal} {Phys. Rev. Lett.}\ }\textbf {\bibinfo
  {volume} {121}},\ \bibinfo {pages} {141101} (\bibinfo {year} {2018})},\
  \Eprint {http://arxiv.org/abs/1712.02240} {arXiv:1712.02240 [astro-ph.CO]}
  \BibitemShut {NoStop}%
%%CITATION = ARXIV:1712.02240;%%
\bibitem [{\citenamefont {Allsman}\ \emph {et~al.}(2001)\citenamefont {Allsman}
  \emph {et~al.}}]{Allsman:2000kg}%
  \BibitemOpen
  \bibfield  {author} {\bibinfo {author} {\bibfnamefont {R.~A.}\ \bibnamefont
  {Allsman}} \emph {et~al.} (\bibinfo {collaboration} {Macho}),\ }\href
  {\doibase 10.1086/319636} {\bibfield  {journal} {\bibinfo  {journal}
  {Astrophys. J.}\ }\textbf {\bibinfo {volume} {550}},\ \bibinfo {pages} {L169}
  (\bibinfo {year} {2001})},\ \Eprint {http://arxiv.org/abs/astro-ph/0011506}
  {arXiv:astro-ph/0011506 [astro-ph]} \BibitemShut {NoStop}%
%%CITATION = ASTRO-PH/0011506;%%
\bibitem [{\citenamefont {Koushiappas}\ and\ \citenamefont
  {Loeb}(2017)}]{Koushiappas:2017chw}%
  \BibitemOpen
  \bibfield  {author} {\bibinfo {author} {\bibfnamefont {S.~M.}\ \bibnamefont
  {Koushiappas}}\ and\ \bibinfo {author} {\bibfnamefont {A.}~\bibnamefont
  {Loeb}},\ }\href {\doibase 10.1103/PhysRevLett.119.041102} {\bibfield
  {journal} {\bibinfo  {journal} {Phys. Rev. Lett.}\ }\textbf {\bibinfo
  {volume} {119}},\ \bibinfo {pages} {041102} (\bibinfo {year} {2017})},\
  \Eprint {http://arxiv.org/abs/1704.01668} {arXiv:1704.01668 [astro-ph.GA]}
  \BibitemShut {NoStop}%
%%CITATION = ARXIV:1704.01668;%%
\bibitem [{\citenamefont {Brandt}(2016)}]{Brandt:2016aco}%
  \BibitemOpen
  \bibfield  {author} {\bibinfo {author} {\bibfnamefont {T.~D.}\ \bibnamefont
  {Brandt}},\ }\href {\doibase 10.3847/2041-8205/824/2/L31} {\bibfield
  {journal} {\bibinfo  {journal} {Astrophys. J.}\ }\textbf {\bibinfo {volume}
  {824}},\ \bibinfo {pages} {L31} (\bibinfo {year} {2016})},\ \Eprint
  {http://arxiv.org/abs/1605.03665} {arXiv:1605.03665 [astro-ph.GA]}
  \BibitemShut {NoStop}%
%%CITATION = ARXIV:1605.03665;%%
\bibitem [{\citenamefont {Monroy-Rodriguez}\ and\ \citenamefont
  {Allen}(2014)}]{Monroy-Rodriguez:2014ula}%
  \BibitemOpen
  \bibfield  {author} {\bibinfo {author} {\bibfnamefont {M.~A.}\ \bibnamefont
  {Monroy-Rodriguez}}\ and\ \bibinfo {author} {\bibfnamefont {C.}~\bibnamefont
  {Allen}},\ }\href {\doibase 10.1088/0004-637X/790/2/159} {\bibfield
  {journal} {\bibinfo  {journal} {Astrophys. J.}\ }\textbf {\bibinfo {volume}
  {790}},\ \bibinfo {pages} {159} (\bibinfo {year} {2014})},\ \Eprint
  {http://arxiv.org/abs/1406.5169} {arXiv:1406.5169 [astro-ph.GA]} \BibitemShut
  {NoStop}%
%%CITATION = ARXIV:1406.5169;%%
\bibitem [{\citenamefont {Ali-Haimoud}\ and\ \citenamefont
  {Kamionkowski}(2017)}]{Ali-Haimoud:2016mbv}%
  \BibitemOpen
  \bibfield  {author} {\bibinfo {author} {\bibfnamefont {Y.}~\bibnamefont
  {Ali-Haimoud}}\ and\ \bibinfo {author} {\bibfnamefont {M.}~\bibnamefont
  {Kamionkowski}},\ }\href {\doibase 10.1103/PhysRevD.95.043534} {\bibfield
  {journal} {\bibinfo  {journal} {Phys. Rev.}\ }\textbf {\bibinfo {volume}
  {D95}},\ \bibinfo {pages} {043534} (\bibinfo {year} {2017})},\ \Eprint
  {http://arxiv.org/abs/1612.05644} {arXiv:1612.05644 [astro-ph.CO]}
  \BibitemShut {NoStop}%
%%CITATION = ARXIV:1612.05644;%%
\bibitem [{\citenamefont {Carr}\ \emph
  {et~al.}(2017{\natexlab{b}})\citenamefont {Carr}, \citenamefont {Raidal},
  \citenamefont {Tenkanen}, \citenamefont {Vaskonen},\ and\ \citenamefont
  {Veermae}}]{Carr:2017jsz}%
  \BibitemOpen
  \bibfield  {author} {\bibinfo {author} {\bibfnamefont {B.}~\bibnamefont
  {Carr}}, \bibinfo {author} {\bibfnamefont {M.}~\bibnamefont {Raidal}},
  \bibinfo {author} {\bibfnamefont {T.}~\bibnamefont {Tenkanen}}, \bibinfo
  {author} {\bibfnamefont {V.}~\bibnamefont {Vaskonen}}, \ and\ \bibinfo
  {author} {\bibfnamefont {H.}~\bibnamefont {Veermae}},\ }\href {\doibase
  10.1103/PhysRevD.96.023514} {\bibfield  {journal} {\bibinfo  {journal} {Phys.
  Rev.}\ }\textbf {\bibinfo {volume} {D96}},\ \bibinfo {pages} {023514}
  (\bibinfo {year} {2017}{\natexlab{b}})},\ \Eprint
  {http://arxiv.org/abs/1705.05567} {arXiv:1705.05567 [astro-ph.CO]}
  \BibitemShut {NoStop}%
%%CITATION = ARXIV:1705.05567;%%
\bibitem [{\citenamefont {Carr}\ \emph {et~al.}(2016)\citenamefont {Carr},
  \citenamefont {Kuhnel},\ and\ \citenamefont {Sandstad}}]{Carr2016}%
  \BibitemOpen
  \bibfield  {author} {\bibinfo {author} {\bibfnamefont {B.}~\bibnamefont
  {Carr}}, \bibinfo {author} {\bibfnamefont {F.}~\bibnamefont {Kuhnel}}, \ and\
  \bibinfo {author} {\bibfnamefont {M.}~\bibnamefont {Sandstad}},\ }\href
  {\doibase 10.1103/PhysRevD.94.083504} {\bibfield  {journal} {\bibinfo
  {journal} {Phys. Rev.}\ }\textbf {\bibinfo {volume} {D94}},\ \bibinfo {pages}
  {083504} (\bibinfo {year} {2016})},\ \Eprint
  {http://arxiv.org/abs/1607.06077} {arXiv:1607.06077 [astro-ph.CO]}
  \BibitemShut {NoStop}%
%%CITATION = ARXIV:1607.06077;%%
\bibitem [{\citenamefont {Bellomo}\ \emph {et~al.}(2018)\citenamefont
  {Bellomo}, \citenamefont {Bernal}, \citenamefont {Raccanelli},\ and\
  \citenamefont {Verde}}]{Bellomo:2017zsr}%
  \BibitemOpen
  \bibfield  {author} {\bibinfo {author} {\bibfnamefont {N.}~\bibnamefont
  {Bellomo}}, \bibinfo {author} {\bibfnamefont {J.~L.}\ \bibnamefont {Bernal}},
  \bibinfo {author} {\bibfnamefont {A.}~\bibnamefont {Raccanelli}}, \ and\
  \bibinfo {author} {\bibfnamefont {L.}~\bibnamefont {Verde}},\ }\href
  {\doibase 10.1088/1475-7516/2018/01/004} {\bibfield  {journal} {\bibinfo
  {journal} {JCAP}\ }\textbf {\bibinfo {volume} {1801}},\ \bibinfo {pages}
  {004} (\bibinfo {year} {2018})},\ \Eprint {http://arxiv.org/abs/1709.07467}
  {arXiv:1709.07467 [astro-ph.CO]} \BibitemShut {NoStop}%
%%CITATION = ARXIV:1709.07467;%%
\bibitem [{\citenamefont {Carr}\ \emph
  {et~al.}(2017{\natexlab{c}})\citenamefont {Carr}, \citenamefont {Tenkanen},\
  and\ \citenamefont {Vaskonen}}]{Carr2017}%
  \BibitemOpen
  \bibfield  {author} {\bibinfo {author} {\bibfnamefont {B.}~\bibnamefont
  {Carr}}, \bibinfo {author} {\bibfnamefont {T.}~\bibnamefont {Tenkanen}}, \
  and\ \bibinfo {author} {\bibfnamefont {V.}~\bibnamefont {Vaskonen}},\ }\href
  {\doibase 10.1103/PhysRevD.96.063507} {\bibfield  {journal} {\bibinfo
  {journal} {Phys. Rev.}\ }\textbf {\bibinfo {volume} {D96}},\ \bibinfo {pages}
  {063507} (\bibinfo {year} {2017}{\natexlab{c}})},\ \Eprint
  {http://arxiv.org/abs/1706.03746} {arXiv:1706.03746 [astro-ph.CO]}
  \BibitemShut {NoStop}%
%%CITATION = ARXIV:1706.03746;%%
\end{thebibliography}

\end{document}